\documentclass[12pt]{elsarticle}
\usepackage{geometry}
\usepackage{graphicx}
\usepackage{array}
\usepackage{multirow}
\usepackage{amsthm}
\usepackage{here}
\usepackage{lscape}
\usepackage{amsmath}
\newtheorem{lem}{\underline{Lemma}} 

\usepackage{amssymb}

\journal{Computational Statistics and Data Analysis}

\begin{document}

\begin{frontmatter}



\title{Flexible estimation in cure survival models using Bayesian P-splines}


\author[a]{Vincent Bremhorst\corref{cor1}}
\ead{vincent.bremhorst@uclouvain.be}
\author[a,b]{Philippe Lambert}
\ead{p.lambert@ulg.ac.be}

\address[a]{Universit\'e catholique de Louvain, Institut de Statistique, Biostatistique et Sciences Actuarielles, Voie du Roman Pays 20, B-1348 Louvain-la-Neuve, Belgium}
\address[b]{Institut des sciences humaines et sociales, M\'ethodes quantitatives en sciences
sociales,Universit\'e de Li\`ege, Belgium, boulevard du Rectorat 7, 
4000 Li\`ege, Belgium}
\cortext[cor1]{Corresponding author}
\begin{abstract}
In the analysis of survival data, it is usually assumed that any unit will experience the event of interest if it is observed for a sufficiently long time. However, it can be explicitly assumed that an unknown proportion of the population under study will never experience the monitored event. The promotion time model, which has a biological motivation, is one of the survival models taking this feature into account. The promotion time model assumes that the failure time of each subject is generated by the minimum of $N$ independent latent event times with a common distribution independent of $N$. An extension which allows the covariates to influence simultaneously the probability of being cured and the latent distribution is presented. The latent distribution is estimated using a flexible Cox proportional hazard model where the logarithm of the baseline hazard function is specified using Bayesian P-splines. Introducing covariates in the latent distribution implies that the population hazard function might not have a proportional hazard structure. However, the use of P-splines provides a smooth estimation of the population hazard ratio over time. The identification issues of the model are discussed and a restricted use of the model when the follow up of the study is not sufficiently long is proposed. The accuracy of our methodology is evaluated through a simulation study and the model is illustrated on data from a Melanoma clinical trial.

\end{abstract}


\end{frontmatter}


\section{Introduction}
\label{Bremhorst::intro}
\noindent
A common hypothesis in the analysis of survival data is that any observed unit will experience the monitored event if it is observed for a sufficient long time. For example, in a cancer clinical trial, one implicitly assumes that all patients will be observed to have a relapse if their follow up is long enough. Hopefully, this is not always a realistic assumption and the consequences of such a wrong hypothesis on the results of the analysis is more and more questioned in the survival literature. \\
Alternatively, one can explicitly acknowledge that an unknown and unidentified proportion of the population under study is cured and will never experience the event of interest. Such models are refered as cure survival models. There are two well known families of cure survival models.
The first one, often refered as the standard mixture cure model, assumes that the population survival function is obtained as a mixture of contributions due to susceptible and cured individuals :

\begin{equation}
S_p(t| \mathbf{x}, \mathbf{z}) = p(\mathbf{x}) S_u(t|\mathbf{z}) + (1 - p(\mathbf{x})),
\end{equation}
where $p(\mathbf{x})$ is the probability of being susceptible and $S_u(t|\mathbf{z})$ is the survival function of the susceptible individuals. This family of cure models was first introduced by Berkson and Gage (1952). They consider $p(\mathbf{x})$ as an unknown constant and $S_u(t |\mathbf{z})$ is related to a parametric model. Farewell (1982 and 1986) extends the model by letting the covariates influence the probability of being susceptible through a logistic regression. Besides that specification for $p(\mathbf{x})$, many authors propose a semiparametric model for the susceptible survival function, see for example Kuk and Chen (1992), Taylor (1995), Peng and Dear (2000), Sy and Taylor (2000), Li and Taylor (2002), Peng (2003), Lu (2010) and Zhang, Peng and Li (2013). Wang, Du and Liang (2012) propose the first completely nonparametric mixture cure model.\\
The second family, often refered as the promotion time (cure) model or as the non-mixture cure model, was developed and studied by Yakovlev and Tsodikov (1996), Tsodikov (1998) and Chen, Ibrahim and Sinha (1999). The promotion time model is motivated using biological mechanisms in the development of cancer. The model argues that each subject is exposed to a number $N \sim P_{ois}(\theta)$ of carcinogenic cells. For each cell, $Y$ is defined as the time necessary for it to yield a detectable cancer mass. The $Y_i's$ are often refered as the latent event times. We assume that the cancer mass in each cell is detected independently from each other and that only one cell needs to be activated for a subject to fail. The latent event times $\{Y_1, ..., Y_N\}$ are independent with a common proper distribution $F(t)$ independent of $N$ and the observed failure time is defined as $T = \min_i\{Y_i\}$. If the subject is not exposed to carcinogenic cells (if $N = 0$), he or she is considered as cured. Using the biological derivation (Yakovlev and Tsodikov (1996) and Chen et al. (1999)) or the mathematical properties (Tsodikov (1998)) of the model, one can show that the population survival function is given by : 

\begin{equation}
\label{Bremhorst:PromTime}
S_p(t) = \exp\left[-\theta F(t)\right] = \exp\left[-\theta (1-S(t))\right]. 
\end{equation}
Note that, since $F(t)$ is a proper cumulative distribution function, the probability of being cured is given by $P(N = 0) = \lim_{t \rightarrow \infty} S_p(t) =  \exp(-\theta)$. \\
When the covariates only influence the probability of being cured, the log-link is usually used on the parameter $\theta$. Many approaches were proposed to specify the latent distribution $F(t)$, see for example Ibrahim, Chen and Sinha (2001), Zeng, Yin and Ibrahim (2006) and Liu and Shen (2009). \\
Since the covariates might jointly influence the probability to be cured and the time necessary for a cell to yield a detectable tumor, a Cox proportional hazard model can be suggested for the latent distribution : $F(t|\mathbf{z}) = 1-S_0(t)^{\exp(\mathbf{z}^T \mathbf{\gamma})}$, where $S_0(t)$ is the baseline survival function. Using these two covariates structures, the population survival function defined in (\ref{Bremhorst:PromTime}) becomes : 

\begin{eqnarray}
S_p(t |\mathbf{x}, \mathbf{z}) &=& \exp \left[-\theta(\mathbf{x})F(t|\mathbf{z})\right] \nonumber \\
& =& \exp \left[-\exp\left(\beta_0 + \mathbf{x}^T \mathbf{\beta}\right)\left(1 - S_0(t)^{\exp(\mathbf{z}^T \mathbf{\gamma})}\right)\right]. 
\label{Bremhorst:modbothcov}
\end{eqnarray}
Model (\ref{Bremhorst:modbothcov}) was already studied, for example by Tsodikov (2002) in a frequentist framework and by Yin and Ibrahim (2005) in a Bayesian framework. Yin and Ibrahim (2005) assume a piecewize exponential distribution for the baseline survival function $S_0(t)$. They use the conditional predictive ordinate criterion to select the appropriate number of intervals. \\
In this paper, we propose a flexible estimation of the baseline distribution that does not require a reference to a model selection criterion. We suggest to specify the baseline log-hazard function as a linear combination of cubic B-splines associated to a predefined (large) number of equidistant knots. A roughness penalty will be used to counterbalance the flexibility of the B-splines (Eilers and Marx, 1996). The use of P-splines provides a smooth estimation of the hazard ratio (of the whole population or of the susceptible population) over time when the model does not have a proportional hazard structure. \\ 
To the best of our knowledge, this is the first time that P-splines are used in a promotion time model. Moreover, since we propose a Bayesian estimation procedure, the confidence bounds of the functional part of the model are directly obtained from the posterior chains. \\
When working with cure survival models, it is usually stressed that the follow up should be sufficiently long. We investigate the identifiability issues when that assumption is not satisfied and propose a restricted use of the model. \\
The remainder of the paper is organised as follows. Section 2 investigates the properties of the proposed model : the hazard ratio and identification issues. A flexible specification of the latent distribution is presented in Section 3. Section 4 is devoted to the Bayesian inference. A simulation study and an application on data from the phase III Melanoma e1684 clinical trial are reported in Section 5 and 6, respectively. A discussion concludes the paper.

\section{Model properties} 
\label{Bremhorst::Properties}
\noindent
Consider the promotion time (cure) model with covariates ($\mathbf{x}$ and $\mathbf{z}$) introduced in Section \ref{Bremhorst::intro}. The population survival function is given by equation (\ref{Bremhorst:modbothcov}) where the cure probability is $\exp\left[-\theta(\mathbf{x})\right]$. Although a Cox model was used to describe the time necessary for a cancerous cell to develop a detectable tumor, the survival function at the patient level is usually not of a proportional hazards type (see below).

\subsection{Hazard ratio issues}
\noindent
The population hazard function $h_p(t| \mathbf{x}, \mathbf{z})$ is defined by 
\begin{eqnarray*}
h_p(t | \mathbf{x}, \mathbf{z}) &=& \frac{-d\left(\log\left[S_p\left(t|\mathbf{x},\mathbf{z}\right]\right)\right)}{dt} \\
 &=& \theta(\mathbf{x})f(t|\mathbf{z}) \\ 
&=& \frac{\exp(\beta_0 + \mathbf{x}^T \mathbf{\beta}) \exp(\mathbf{z}^T \mathbf{\gamma}) f_0(t) S_0(t) ^ {\exp(\mathbf{z}^T \mathbf{\gamma})} }{S_0(t)},
\end{eqnarray*}
where $f_0(t)$ is the baseline density function. \\
The population hazard ratio comparing groups 1 and 2 is given by : 
\begin{eqnarray*}
HR_p &=& \frac{h_p(t | \mathbf{x}_1, \mathbf{z}_1)}{h_p(t | \mathbf{x}_2, \mathbf{z}_2)} \\
& =& \exp\left((\mathbf{x}^T_1 - \mathbf{x}^T_2) \mathbf{\beta}\right) \exp\left((\mathbf{z}^T_1 - \mathbf{z}^T_2) \mathbf{\gamma}\right)S_0(t) ^{\left(\exp(\mathbf{z}^T_1 \mathbf{\gamma}) - \exp(\mathbf{z}^T_2 \mathbf{\gamma})\right)}.
\end{eqnarray*}
Thus, the population hazard ratio $HR_p$ remains constant over time if $\exp(\mathbf{z}^T_1 \mathbf{\gamma}) - \exp(\mathbf{z}^T_2 \mathbf{\gamma})$ = 0, i.e. if $\mathbf{z}_1 = \mathbf{z}_2.$ In practice, this means that the population hazard ratio is constant provided that the contrasted groups share the same values for the covariates affecting the development of cancerous cells in the biological model. \\
As mentioned in Chen et al. (1999), the hazard function of the susceptible individuals, given in (\ref{Bremhorst :: Hazard function susceptible}), does not have a proportional hazard structure. Indeed, one can show that the susceptible survival function is given by : 

\begin{eqnarray*}
S_u(t|\mathbf{x}, \mathbf{z}) &=& P(T > t | N \ge 1, \mathbf{x}, \mathbf{z}) \\ 
&=& \frac{S_p(t|\mathbf{x}, \mathbf{z}) - \exp \left[-\theta(\mathbf{x})\right]}{1-\exp\left[-\theta(\mathbf{x})\right]},
\end{eqnarray*}
and the susceptible hazard function is given by :

\begin{eqnarray}
\label{Bremhorst :: Hazard function susceptible}
h_u(t |\mathbf{x}, \mathbf{z}) &=& \frac{-d\left(\log\left[S_u\left(t|\mathbf{x},\mathbf{z}\right)\right]\right)}{dt} \nonumber\\ 
&=&\frac{S_p(t|\mathbf{x}, \mathbf{z})}{S_p(t|\mathbf{x}, \mathbf{z}) - \exp\left[-\theta(\mathbf{x})\right]} h_p(t | \mathbf{x}, \mathbf{z}) \\ 
&=& \frac{h_p(t | \mathbf{x}, \mathbf{z})}{P(T < +\infty | T > t, \mathbf{x}, \mathbf{z})}. \nonumber
\end{eqnarray}
It is straightforward to see that expression (\ref{Bremhorst :: Hazard function susceptible}) does not have a proportional hazard structure.

\subsection{Identification issues}
\label{Bremhorst::Identifiability}
\noindent
Theoretically, the follow up of a study is said to be sufficiently long if the largest censoring time is greater than the largest failure time, i.e. if the follow-up time of any susceptible unit was sufficiently long to observe its failure. In practice, if a plateau is present in the right tail of the estimated population survival function (for example, in the Kaplan Meier estimated curve), the sufficient follow up assumption seems to be reasonable.\\
\begin{lem}
\begin{enumerate}
\item[ ]
\item[ ] Assumptions : 
\item[A1 ] The vector $\mathbf{z}$ of covariates does not include an intercept.
\item[A2 ] $\mathbf{X}^T\mathbf{X}$ and $\mathbf{Z}^T\mathbf{Z}$ are full rank matrices, where $\mathbf{X}$ and $\mathbf{Z}$ are the design matrices corresponding to covariate vectors $\mathbf{x}$ and $\mathbf{z}$, respectively.  
\item[A3 ] The baseline cumulative distribution function $F_0(t) = 1-S_0(t)$ is proper.
\item[ ] Under A1, A2 and A3, we have : 
\item[1) ] If the follow up of the study is sufficiently long, then model (\ref{Bremhorst:modbothcov}) is identifiable.
\item[2) ] If the follow up of the study is not sufficiently long and if vectors $\mathbf{x}$ and $\mathbf{z}$ do not share some components, then only the estimations of the covariate effects on the cure probability and on failure time for a cancerous cell are identifiable. 
\end{enumerate}
\end{lem}
\noindent
The proof of this lemma is in appendix: it is based on the proof proposed by Liu and Shen (2009) showing the identifiability of the promotion time model when the covariates only influence the probability to be cured. 

\section{Flexible specification of the baseline distribution}
\label{Bremhorst::FlexibleSpecification}
\noindent
We assume some familiarity with P-splines from the reader. If not, information about P-splines can be found in Eilers and Marx (1996) and in Lang and Brezger (2004).\\
In order to estimate the baseline survival function $S_0(t)$ in (\ref{Bremhorst:modbothcov}), we suggest to write the baseline log-hazard as a linear combination of cubic B-splines:
\begin{eqnarray}
\label{Bremhorst::basehazard}
h_0(t) &=& \exp\left(\sum_{k=1}^K b_k(t)\phi_k\right),
\end{eqnarray}
where $\{b_k(.), k=1, ... , K\}$ denotes the cubic B-splines basis associated to a predefined number of equidistant knots on $[0, t_{Rcens}]$, where $t_{Rcens}$ is the upper bound of the follow up. \\
To ensure enough flexibility, Eilers and Marx (1996) suggest to choose a large number of B-splines and to counterbalance the flexibility by adding to the log-likelihood a roughness penalty based on finite differences of adjacent B-spline parameters : $\tau \sum_k (\Delta^r \phi_k)^2 =\tau \mathbf{\phi}^T\mathbf{D}^T\mathbf{D}\mathbf{\phi}$, where $\tau$ is the penalty parameter and $\mathbf{D}$ is the $r^{th}$ difference penalty matrix. For example, when a third order penalty is specified, the matrix $\mathbf{D}$ is defined as :
\[
\mathbf{D} = \left[
\begin{array}{ccccccc}
   1 & -3 & 3 & -1 & 0 & ... & 0 \\
   0 & 1 & -3 & 3 & -1 & ... & 0 \\
   \vdots & \vdots & \ddots & \ddots & \ddots & \ddots & \vdots \\
   0 & 0  & ... & 1 & -3 & 3 & -1 \\
\end{array}
\right].
\]
P-splines were already used in many different contexts, see for example Eilers and Marx (1996) and Eilers (2007) in a frequentist framework and Lang and Brezger (2004), Lambert and Eilers (2005),  Lambert (2007, 2013) and Cetinyurek and Lambert (2011) in a Bayesian framework. As mentioned in all these references, if $K$ is chosen large enough (between 10 and 20, say), no model selection criterion is needed since all the $K$'s give similar results.\\ 
Knowing the relation between the survival function and the hazard function and using (\ref{Bremhorst::basehazard}) as expression for the hazard function, the baseline survival function $S_0(t)$ is specified as :

\begin{eqnarray}
\label{Bremhorst:baselinesurvival}
S_0(t) &=& \exp\left(-\int_0^t \exp\left[\sum_{k=1}^K b_k(u) \phi_k\right] du\right).
\end{eqnarray}
The integral in (\ref{Bremhorst:baselinesurvival}) has no analytic form and needs to be evaluated numerically. Knowing that our observations are contained in the interval $[0, t_{Rcens}]$, we partition $[0, t_{Rcens}]$ into $J$ (300, say) small bins (of equal width, for simplicity) $J_j = [\tau_{j-1}, \tau_j]$ where $0 = \tau_0 < \tau_1 < ... < \tau_J    = t_{Rcens}$. Let $u_j$ and $\delta_j$ denote the midpoint and the width of $J_j$, respectively. Then, using the rectangle method, (\ref{Bremhorst:baselinesurvival}) can be approximated by :
\begin{eqnarray}
\label{Bremhorst::approxbaselinesurvival}
S_0(t) &\approx& \exp\left( - \sum_{j=1}^{j(t)} \exp\left[\sum_{k=1}^K b_k(u_j) \phi_k\right] \delta_j \right),
\end{eqnarray}
where $j(t)$ indexes the interval containing $t$.\\
For identifiability purpose (see Section \ref{Bremhorst::Identifiability}), we fix the last spline parameter $\phi_K$ to a large enough value (10, say). In this way, we force the estimated baseline survival function $\hat{S}_0(.)$ to be 0 at the end of the follow up.

\section{Bayesian inference}
\subsection{Likelihood}
\noindent
For the $i^{th}$ subject under study, we observe the failure or the censoring time $t_i$, the event indicator $\nu_i$ and two sets of covariates $\mathbf{x}_i$ and $\mathbf{z}_i$. We denote these observable variables by $\mathbf{\mathbb{D}}_i = (t_i, \nu_i, \mathbf{x}_i, \mathbf{z}_i)$. The set of parameters specific to the chosen model is written as $\mathbf{\Phi}$. Then, the data likelihood is given by : 
\[
L(\mathbf{\Phi} | \mathbf{\mathbb{D}}) = \prod_{i=1}^I h_p(t_i)^{\nu_i} S_p(t_i). 
\] 
\subsection{Bayesian Model}
\label{Bremhorst::Model}
\noindent
In a Bayesian setting, the roughness penalty is translated into a prior distribution for the spline parameters (Lang and Brezger, 2004):
\[
\pi(\mathbf{\phi} | \tau) \propto \tau^{\frac{K}{2}} \exp\left( -\frac{\tau}{2} \mathbf{\phi}^T\mathbf{P}\mathbf{\phi}\right),
\]
where $\mathbf{P} = \mathbf{D}^T\mathbf{D} + \epsilon \mathbf{I}_K$ is a full rank matrix for some small quantity $\epsilon$ ($10^{-6}$, say). In other words, a normal distribution with mean 0 and variance-covariance matrix $\mathbf{P}^{-1}$ for the spline parameters is considered. As suggested by Jullion and Lambert (2007), we take a robust specification for the roughness penalty prior distribution : 
\begin{eqnarray*}
\tau | \delta &\sim& G(\frac{\nu}{2}, \frac{\nu \delta}{2}), \\
\delta &\sim & G(a_{\delta}, b_{\delta}),
\end{eqnarray*}
where $G(a,b)$ denotes a Gamma distribution with mean $\frac{a}{b}$ and variance $\frac{a}{b^2}$. \\
They showed that if a small value is chosen for $a_{\delta}$ and $b_{\delta}$ ($10^{-4}$, say), then the choice of $\nu$ (here, set equal to $2$) does not affect the shape of the estimated curve. If prior knowledge (such as monotonicity) is available about the baseline hazard, it can be expressed during the prior elicitation for the spline parameters. If nothing is known a priori about the covariate effects, a large variance normal prior distribution can be used for all the regression parameters. \\
Using Bayes' theorem, the joint posterior distribution is given by : 
\begin{eqnarray}
\label{Bremhorst::Posterior}
\pi(\mathbf{\Phi}| \mathbf{\mathbb{D}})& \propto& L(\mathbf{\Phi} | \mathbf{\mathbb{D}}) \pi(\mathbf{\phi} | \tau) \pi(\tau| \delta) \pi(\delta) \pi(\beta_0,\mathbf{\beta}) \pi(\mathbf{\gamma}).
\end{eqnarray}
Given that all the prior distributions are proper, the posterior distribution $\pi(\mathbf{\Phi}| \mathbf{\mathbb{D}})$ is proper. Only the conditional posterior of $\tau$ and $\delta$ belong to known families of distributions : 
\begin{eqnarray*}
\tau | \mathbf{\phi},\delta, \mathbf{\mathbb{D}} &\sim& G\left(\frac{\nu + K}{2}, \frac{\nu\delta + \mathbf{\phi}^T \mathbf{P} \mathbf{\phi}}{2}\right), \\
\delta | \tau, \mathbf{\mathbb{D}} &\sim & G(a_{\delta} + \frac{\nu}{2}, b_{\delta} + \frac{\nu \tau}{2})  .
\end{eqnarray*}

\subsection{Posterior sample using MCMC}
\label{PosteriorMCMC}
\noindent
A Metropolis step will be used to sample the other conditional distributions. As shown by Lambert (2007), the mixing of the chains can be improved by applying the Metropolis algorithm on a reparametrized posterior distribution. An adequate reparametrization can be suggested by a frequentist estimation of the correlation structure of the spline parameters. To reach that goal, one could use a nonlinear optimizer to compute the mode of the joint posterior distribution in (\ref{Bremhorst::Posterior}), for a fixed value of $\delta$. At convergence, the hessian matrix can be used to assess the posterior correlation between the parameters and to suggest a reparametrization yielding less dependent components (see Lambert, 2007, for more details). \\
Let $\tilde{\beta} = (\beta_0, \mathbf{\beta})$ and $(\mathbf{\phi}^{(0)}, \tau^{(0)}, \delta^{(0)}, \tilde{\mathbf{\beta}}^{(0)}, \mathbf{\gamma}^{(0)})$ be the initial values of the chain selected, for example, using the optimization step described above. 
The MCMC algorithm consists in five main steps to sample the parameters from the posterior. Iteration $m$ proceeds as follows : 
\begin{enumerate}
\item[$\bullet$] Draw $\mathbf{\phi}^{(m)}$ from $\pi(\mathbf{\phi} | \tau^{(m-1)}, \delta^{(m-1)}, \tilde{\mathbf{\beta}}^{(m-1)}, \mathbf{\gamma}^{(m-1)})$ using univariate Metropolis steps (along directions suggested by the reparametrization) ; 
\item[$\bullet$] Draw $\tau^{(m)}$ from $G\left(\frac{\nu + K}{2}, \frac{\nu\delta^{(m-1)} + \mathbf{\phi}^{(m)^{T}} \mathbf{P} \mathbf{\phi}^{(m)}}{2}\right)$ in a Gibbs step ; 
\item[$\bullet$] Draw $\delta^{(m)}$ from $G(a_{\delta} + \frac{\nu}{2}, b_{\delta} + \frac{\nu \tau^{(m)}}{2})$ in a Gibbs step ; 
\item[$\bullet$] Draw $\tilde{\mathbf{\beta}}^{(m)}$ from $\pi(\tilde{\mathbf{\beta}} |\mathbf{\phi}^{(m)}, \tau^{(m)}, \delta^{(m)}, \mathbf{\gamma}^{(m-1)})$ using univariate Metropolis steps ; 
\item[$\bullet$] Draw $\mathbf{\gamma}^{(m)}$ from $\pi(\mathbf{\gamma} |\mathbf{\phi}^{(m)}, \tau^{(m)}, \delta^{(m)}, \tilde{\mathbf{\beta}}^{(m)})$  using univariate Metropolis steps ;
\end{enumerate}
\noindent
The variances of the proposal distributions in the Metropolis steps are tuned automatically using the adaptive procedure proposed by Haario, Saksman and Tamminen (2001) during the burnin to achieve the targeted optimal acceptance rate (Gelman, Roberts and Gilks, 1996 and Roberts and Rosenthal, 2001).
\section{Simulation study}
\label{Bremhorst:Simulation}
\subsection{Sufficiently long follow up}
\label{Bremhorst::Sufficientlylongfollowup}
\begin{figure}[H] 
\begin{center}
\begin{tabular}{cc}
\includegraphics[width=5.35cm,height=5.35cm]{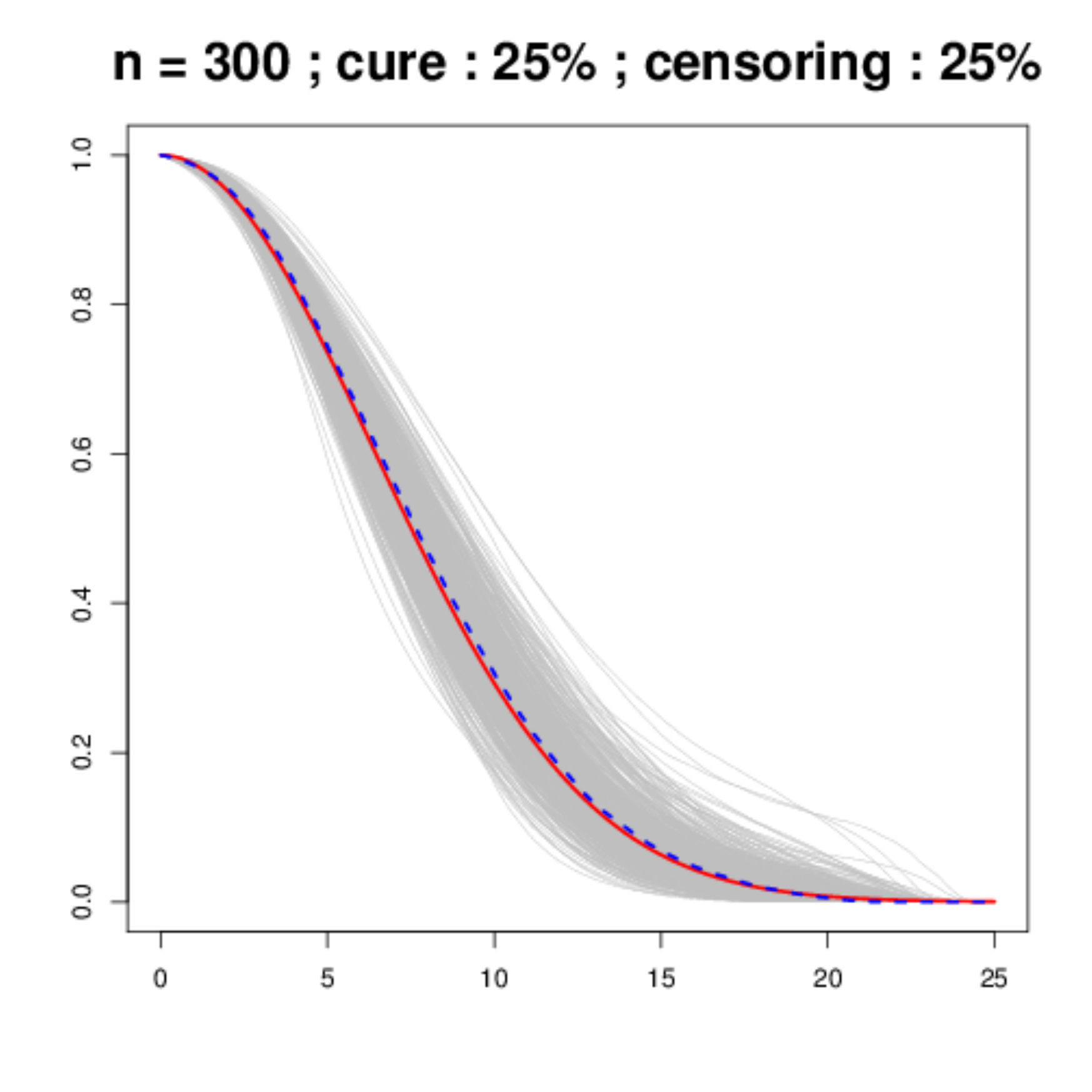} & 
\includegraphics[width=5.35cm,height=5.35cm]{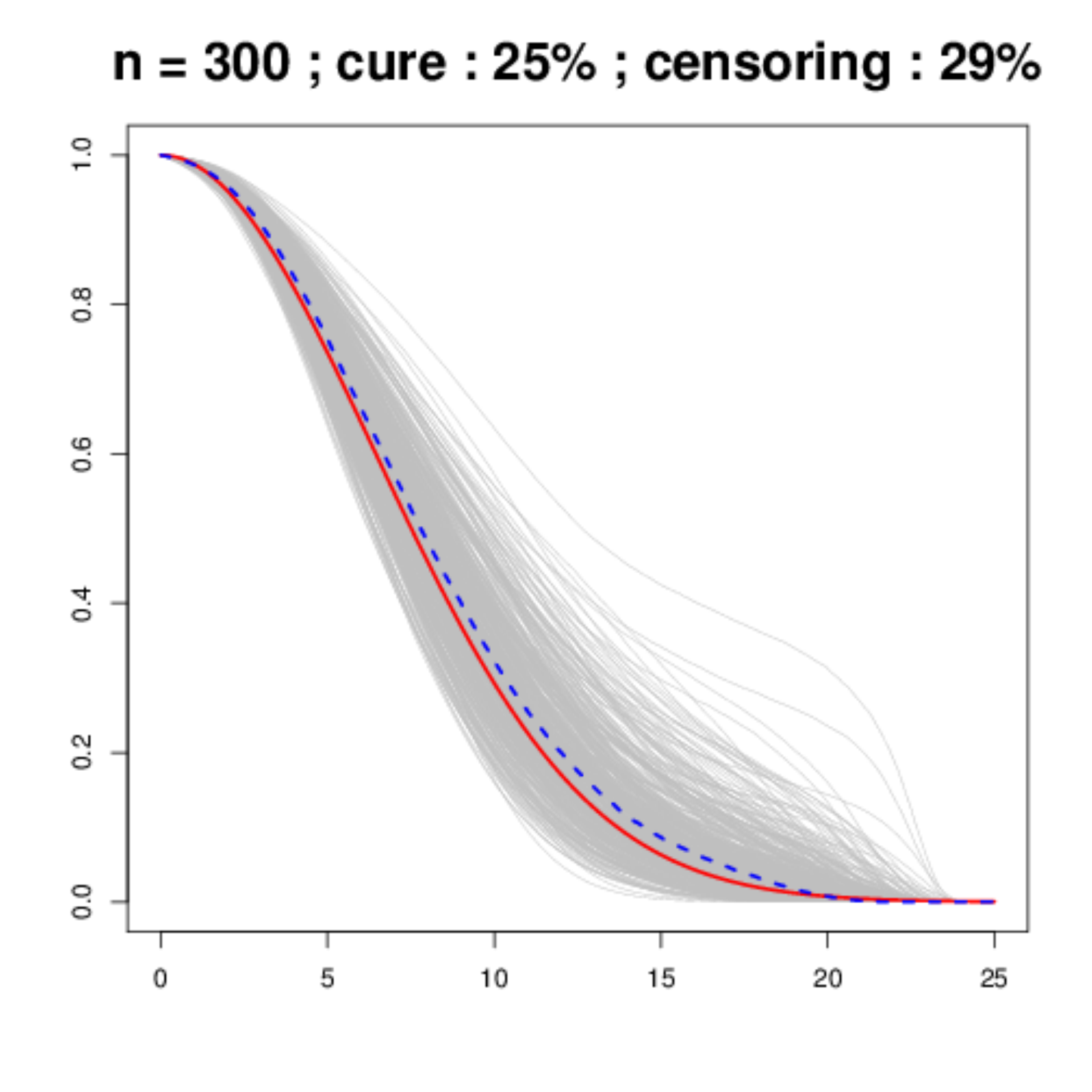}  \\
\includegraphics[width=5.35cm,height=5.35cm]{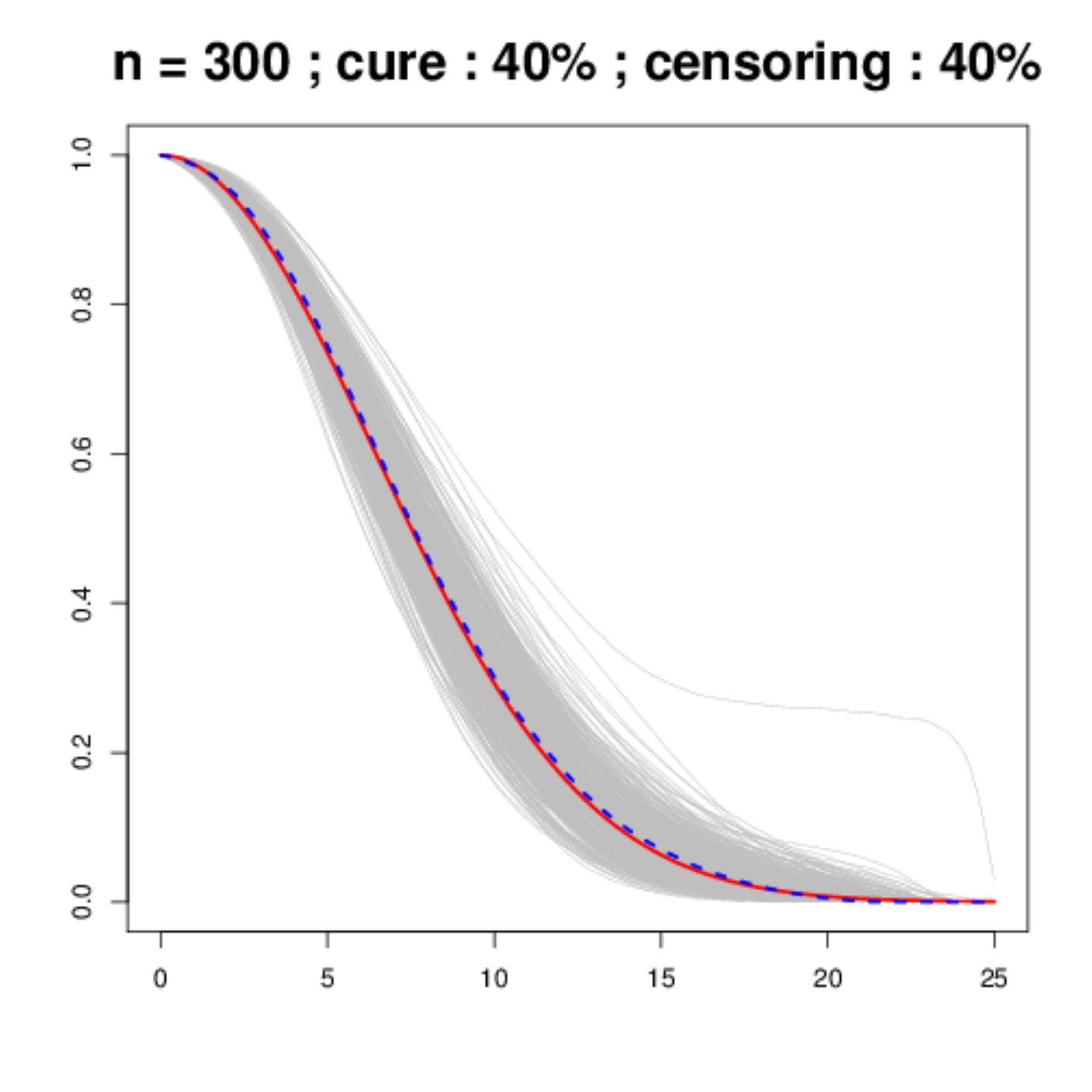} & 
\includegraphics[width=5.35cm,height=5.35cm]{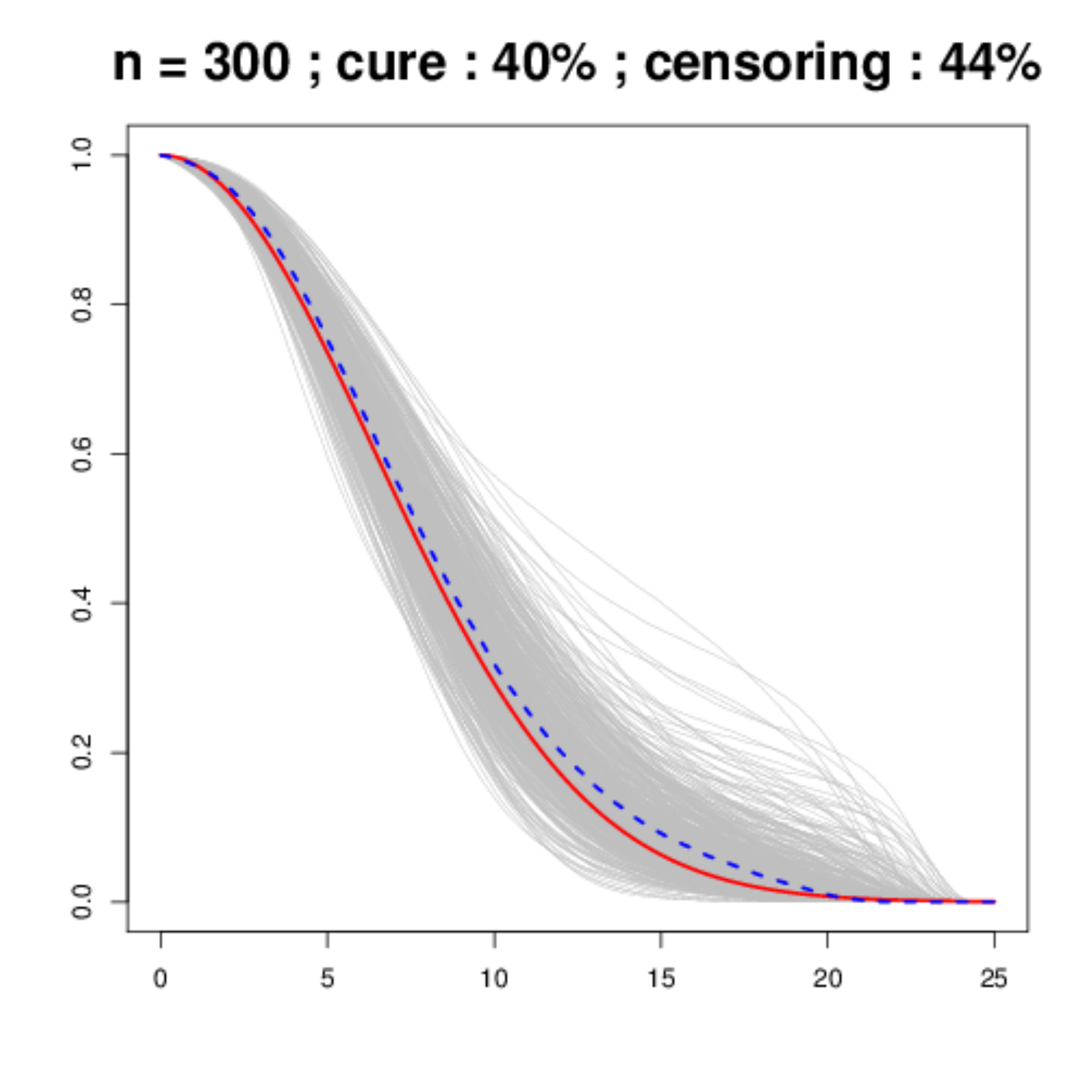}  \\
\end{tabular}
\caption{\footnotesize{Simulation results when the follow up is sufficiently long : estimation of the baseline distribution $S_0(t)$ for $S=500$ replications (one grey curve per data set) and sample size $n=300$. Each row refers to a percentage of cured individuals (row 1 : 25\%, row 2 : 40\%) with different global right censoring rates (left : setting 1 ; right : setting 2). The solid line corresponds to the true function and the dashed line is the pointwize median of the 500 estimated curves.}}
\label{Bremhorst::baseline300}
\end{center}
\end{figure}

\begin{figure}[H] 
\begin{center}
\begin{tabular}{cc}
\includegraphics[width=5.35cm,height=5.35cm]{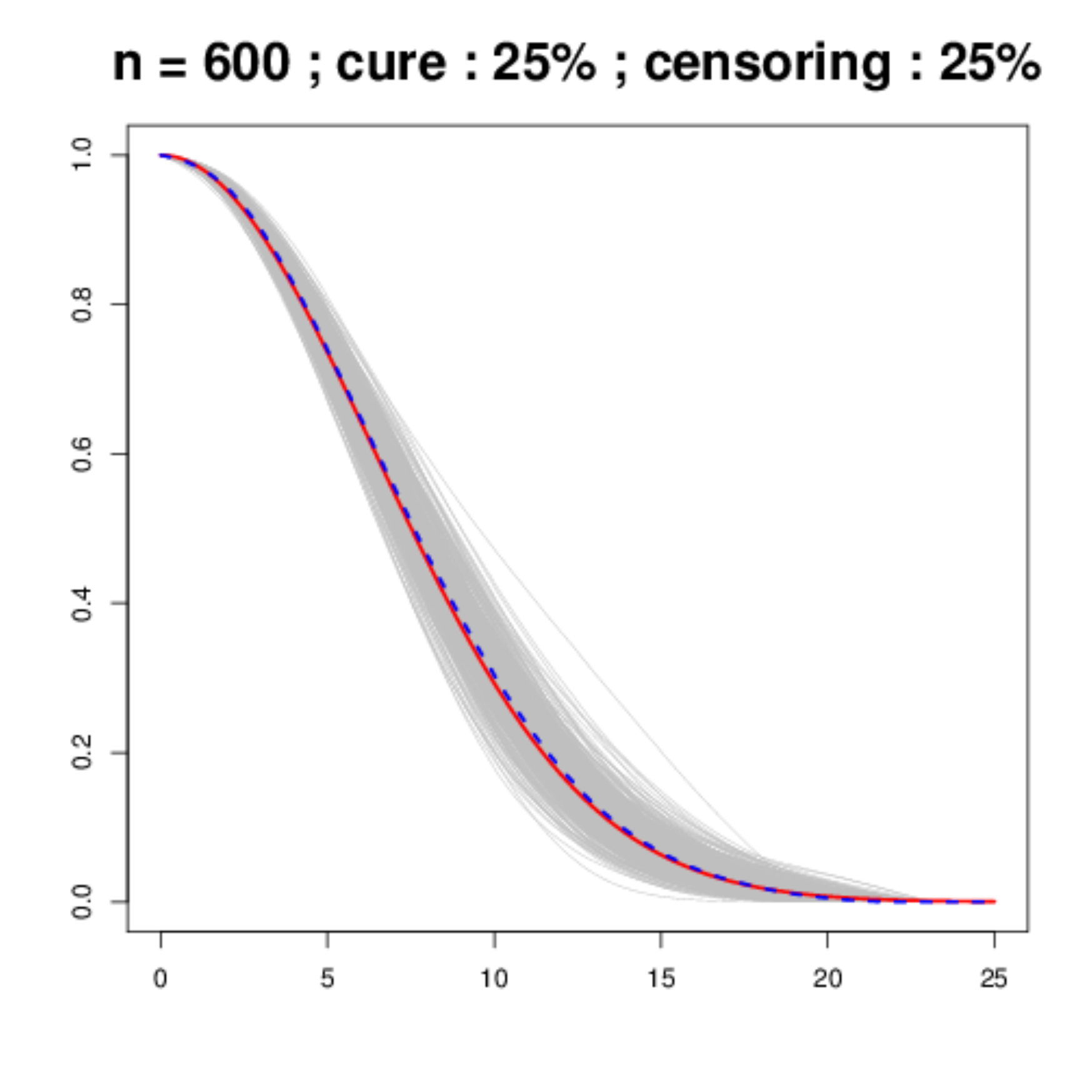} & 
\includegraphics[width=5.35cm,height=5.35cm]{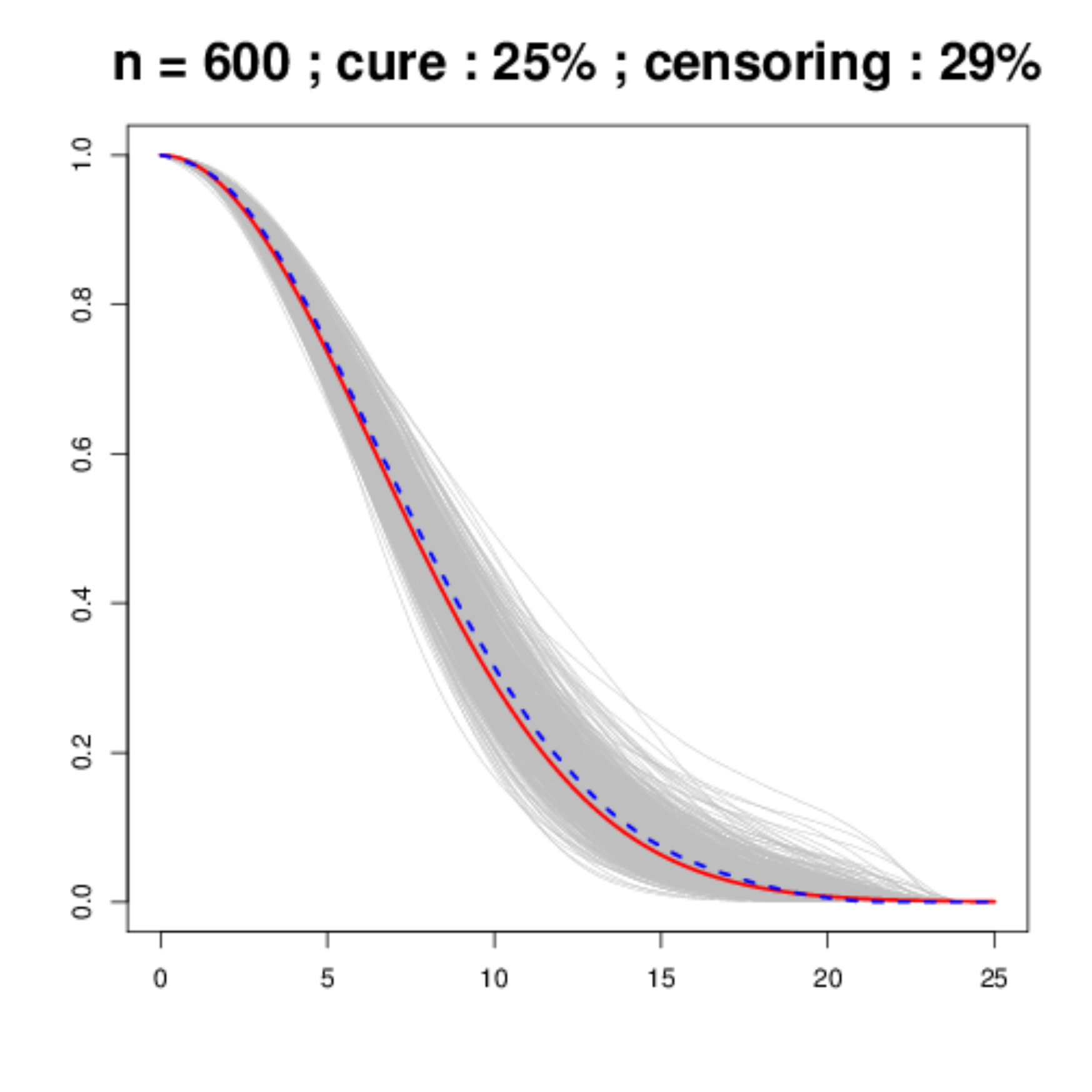}  \\
\includegraphics[width=5.35cm,height=5.35cm]{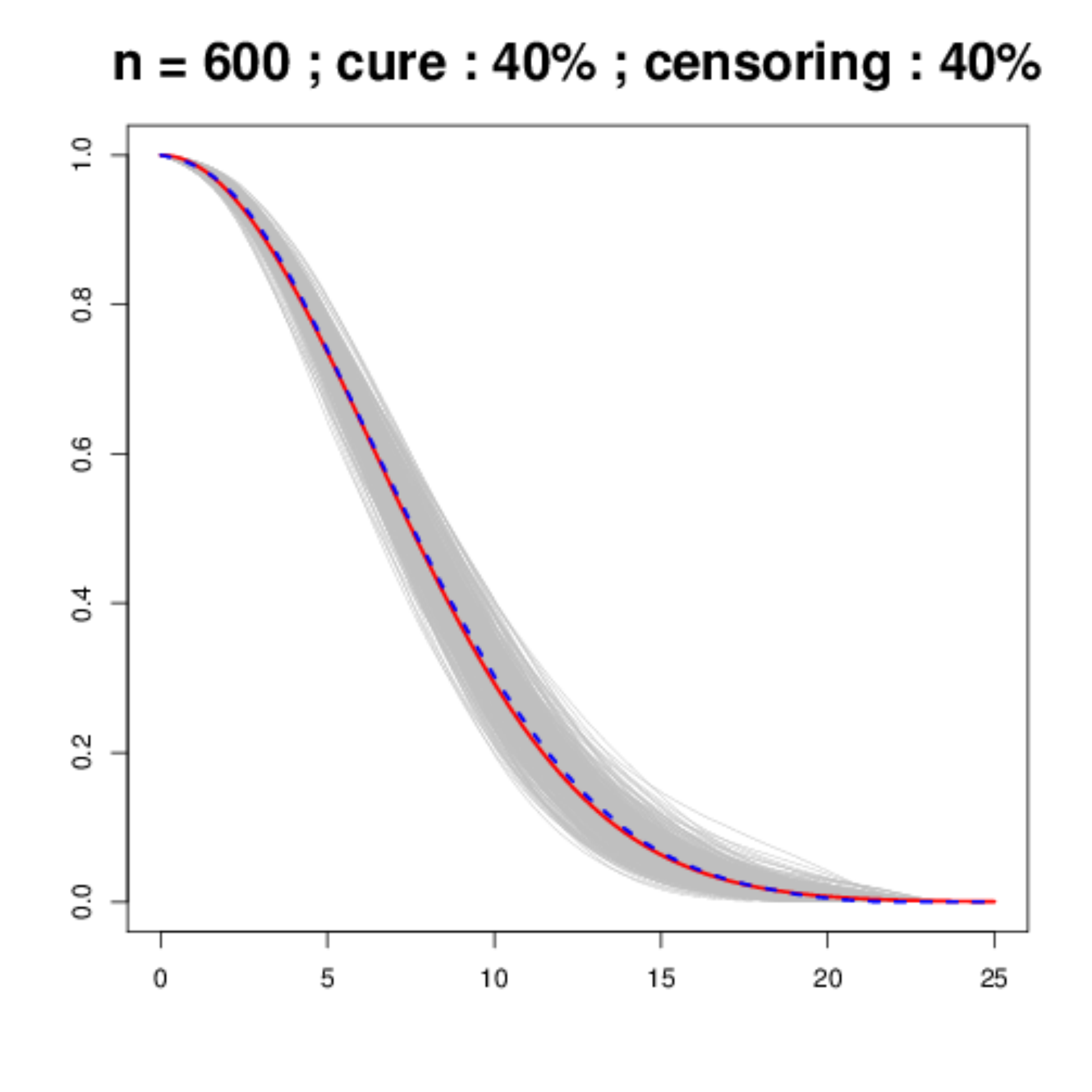} & 
\includegraphics[width=5.35cm,height=5.35cm]{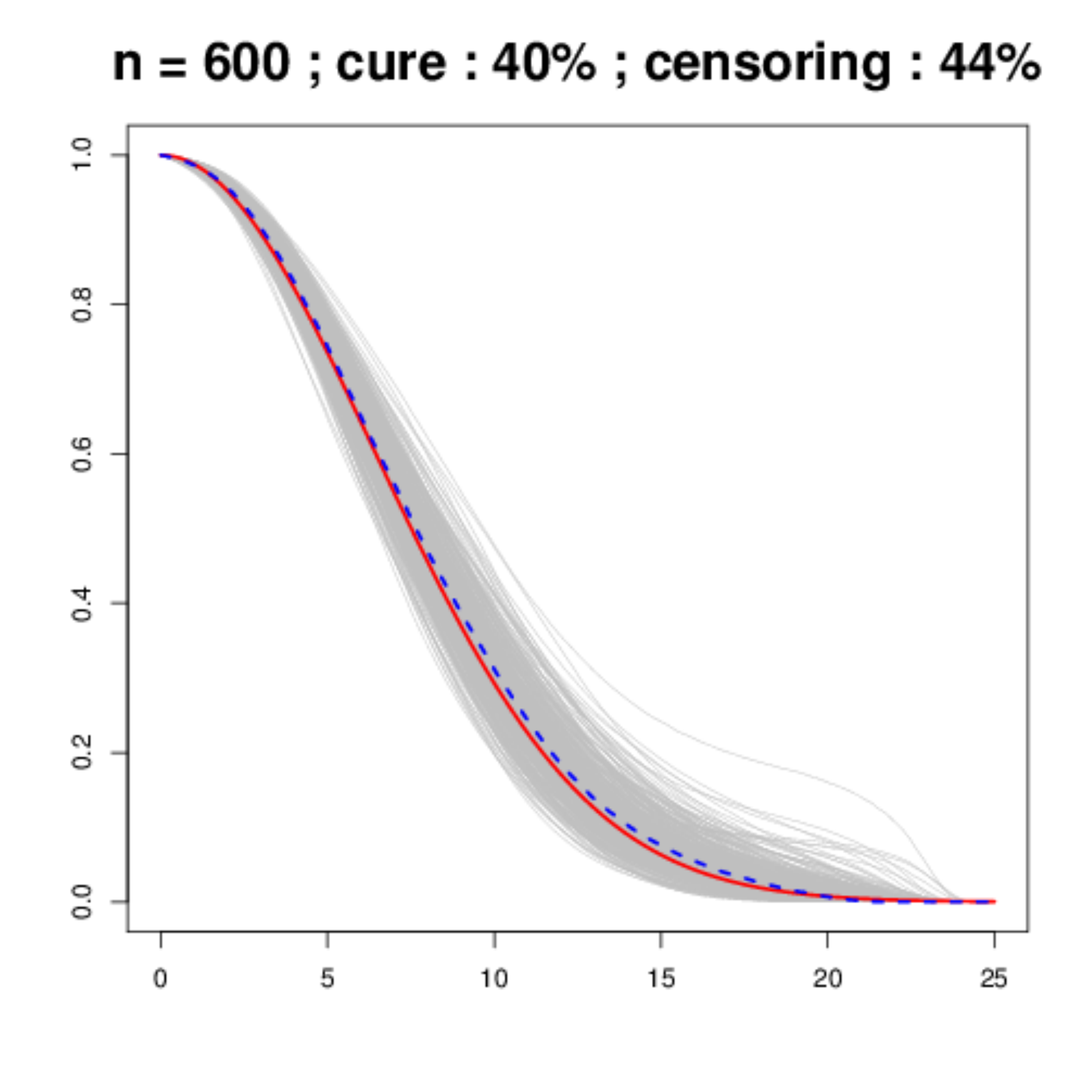}  \\
\end{tabular}
\caption{\footnotesize{Simulation results when the follow up is sufficiently long : estimation of the baseline distribution $S_0(t)$ for $S=500$ replications (one grey curve per data set) and sample size $n=600$. Each row refers to a percentage of cured individuals (row 1 : 25\%, row 2 : 40\%) with different global right censoring rates (left : setting 1 ; right : setting 2). The solid line corresponds to the true function and the dashed line is the pointwize median of the 500 estimated curves.}}
\label{Bremhorst::baseline600}
\end{center}
\end{figure}

\begin{figure}[H] 
\begin{center}
\begin{tabular}{cc}
\includegraphics[width=5.35cm,height=5.35cm]{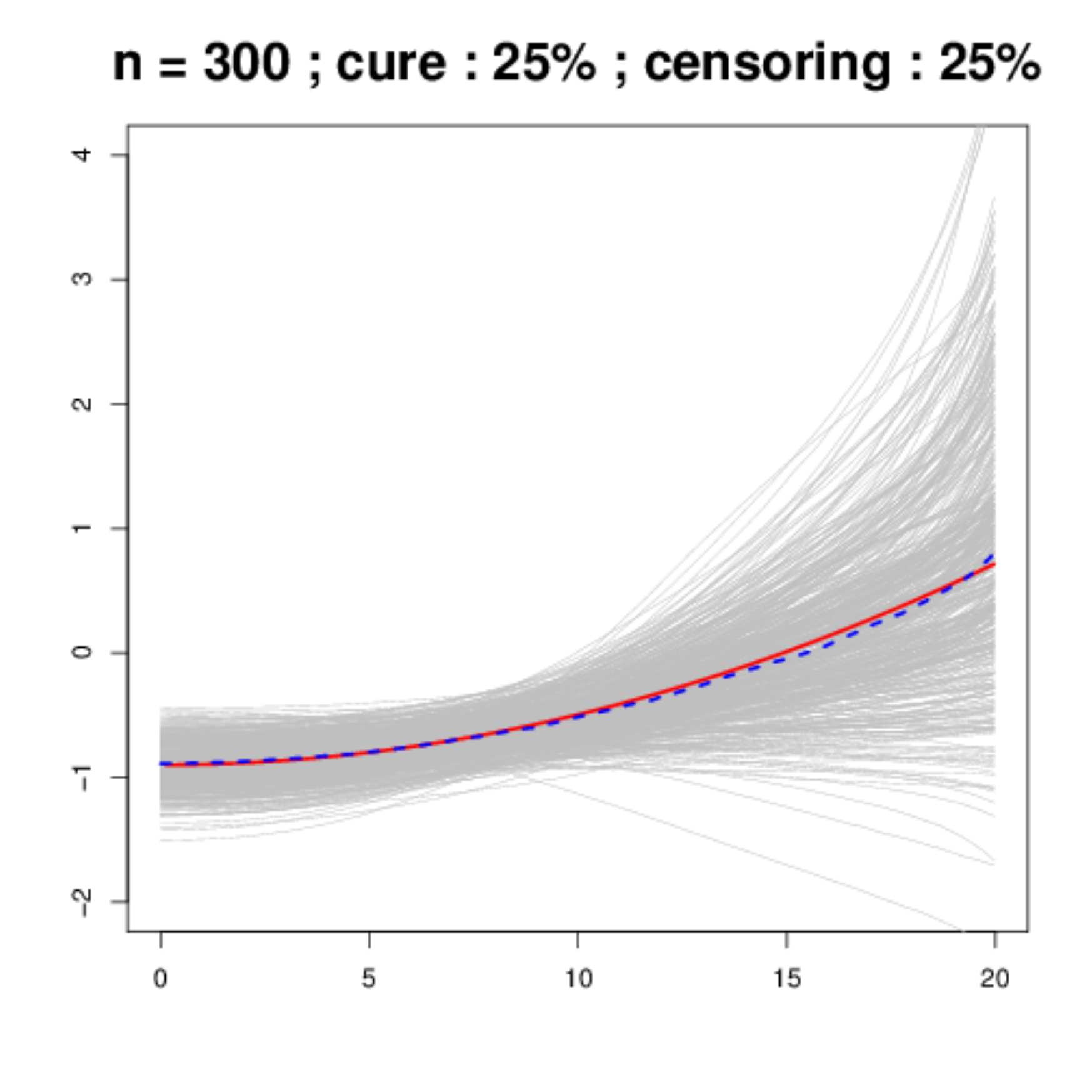} &
 \includegraphics[width=5.35cm,height=5.35cm]{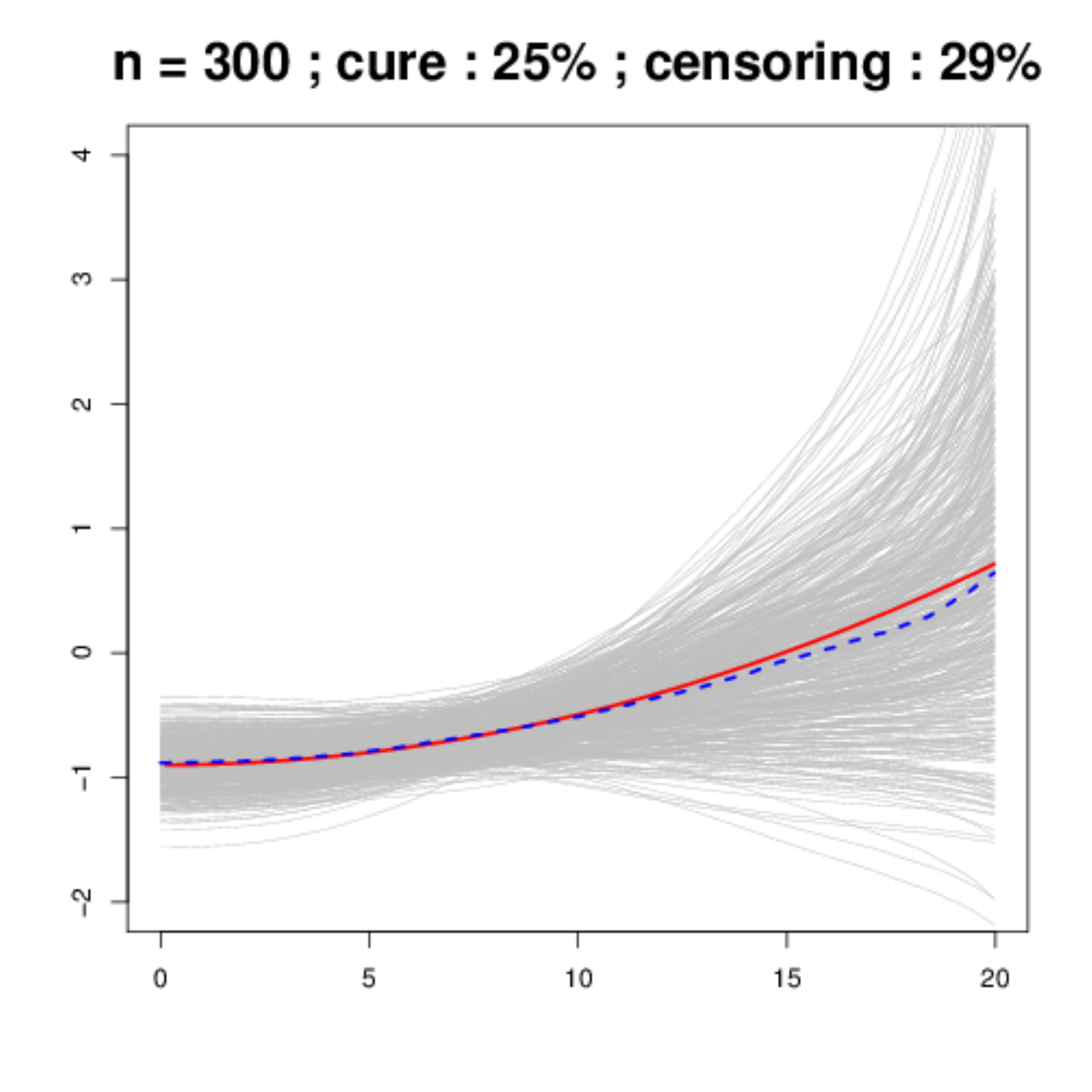}  \\
\includegraphics[width=5.35cm,height=5.35cm]{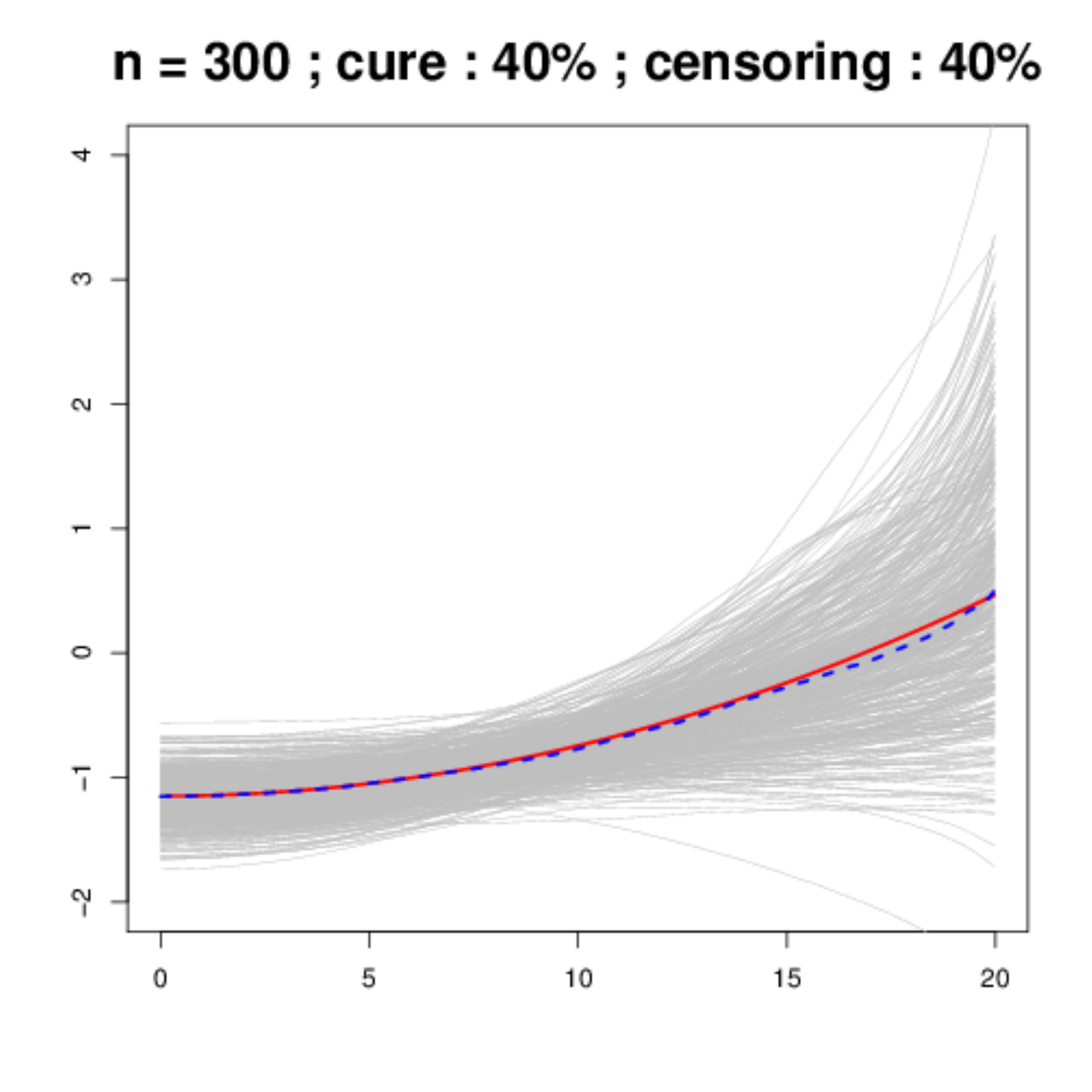} & 
\includegraphics[width=5.35cm,height=5.35cm]{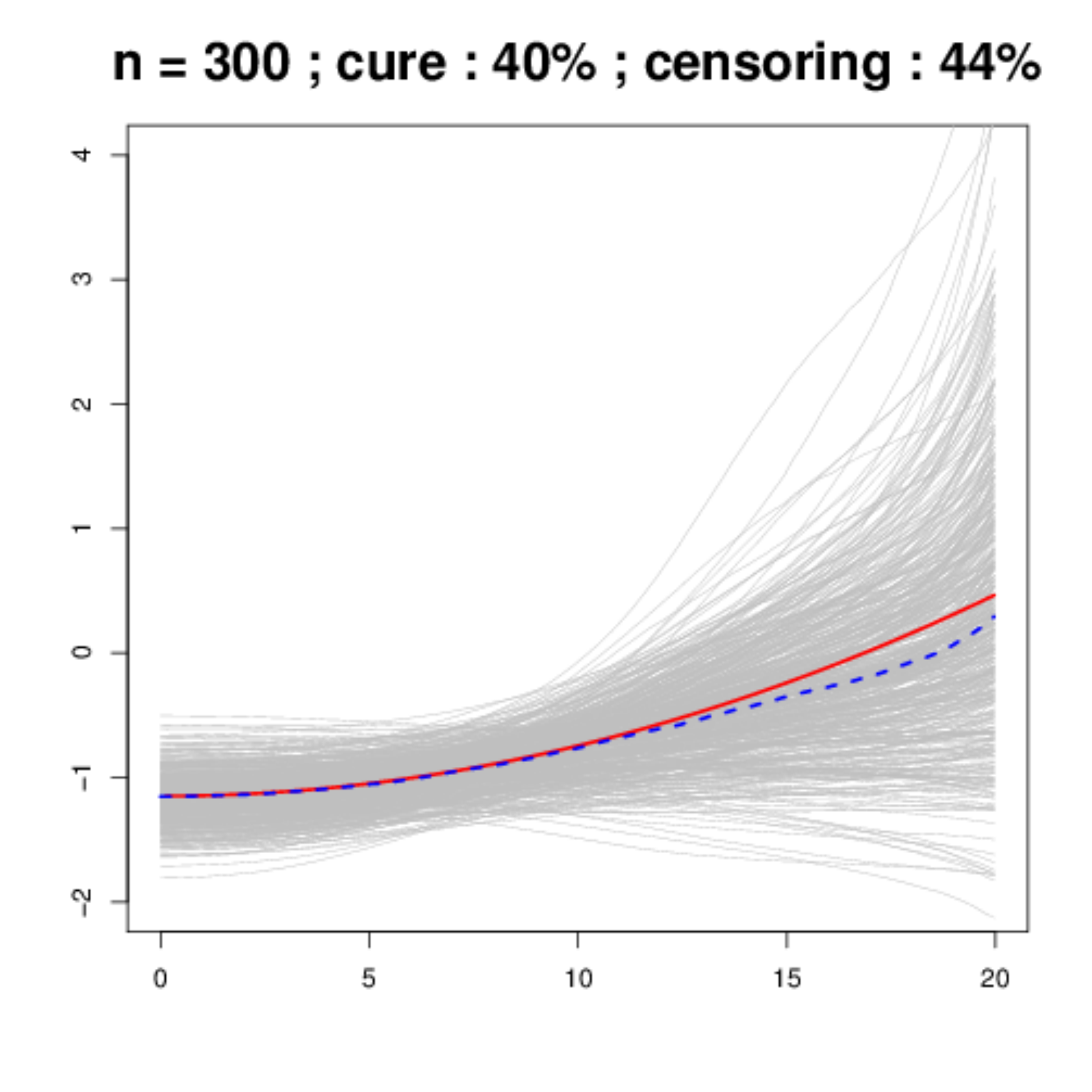}  
\end{tabular}
\caption{\footnotesize{Simulation results when the follow up is sufficiently long : estimation of the population log-hazard ratio $\log(HR_p(t))$ for $S=500$ replications (one grey curve per data set) and sample size $n=300$. Each row refers to a percentage of cured individuals (row 1 : 25\%, row 2 : 40\%) with different global right censoring rates (left : setting 1 ; right : setting 2). The solid line corresponds to the true function and the dashed line is the pointwize median of the 500 estimated curves. The hazard ratio is obtained by contrasting the groups induced by the binary covariate (for a median value of the continuous covariate.)}}
\label{Bremhorst::PopHazard300}
\end{center}
\end{figure}

\begin{figure}[H] 
\begin{center}
\begin{tabular}{cc}
\includegraphics[width=5.35cm,height=5.35cm]{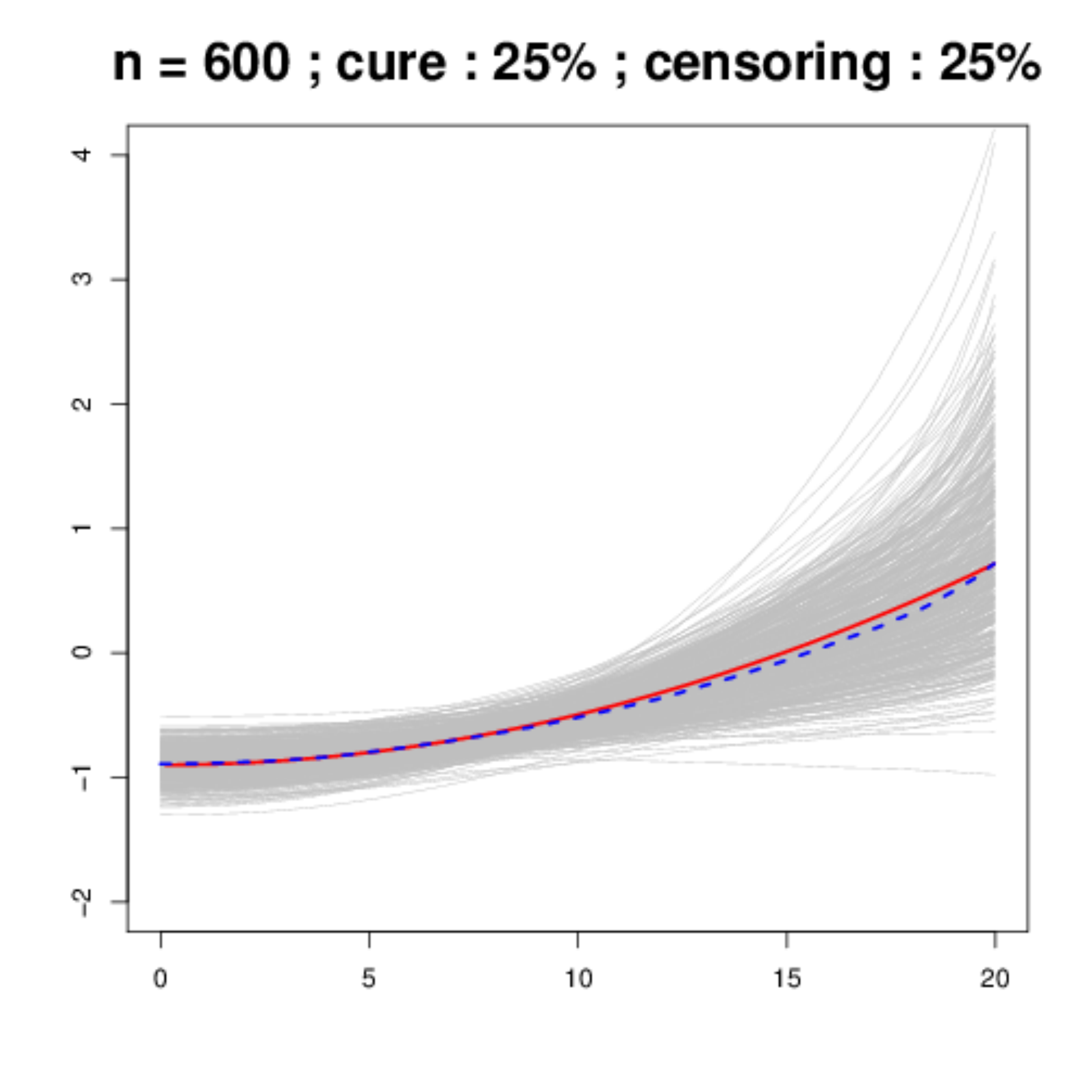} &
 \includegraphics[width=5.35cm,height=5.35cm]{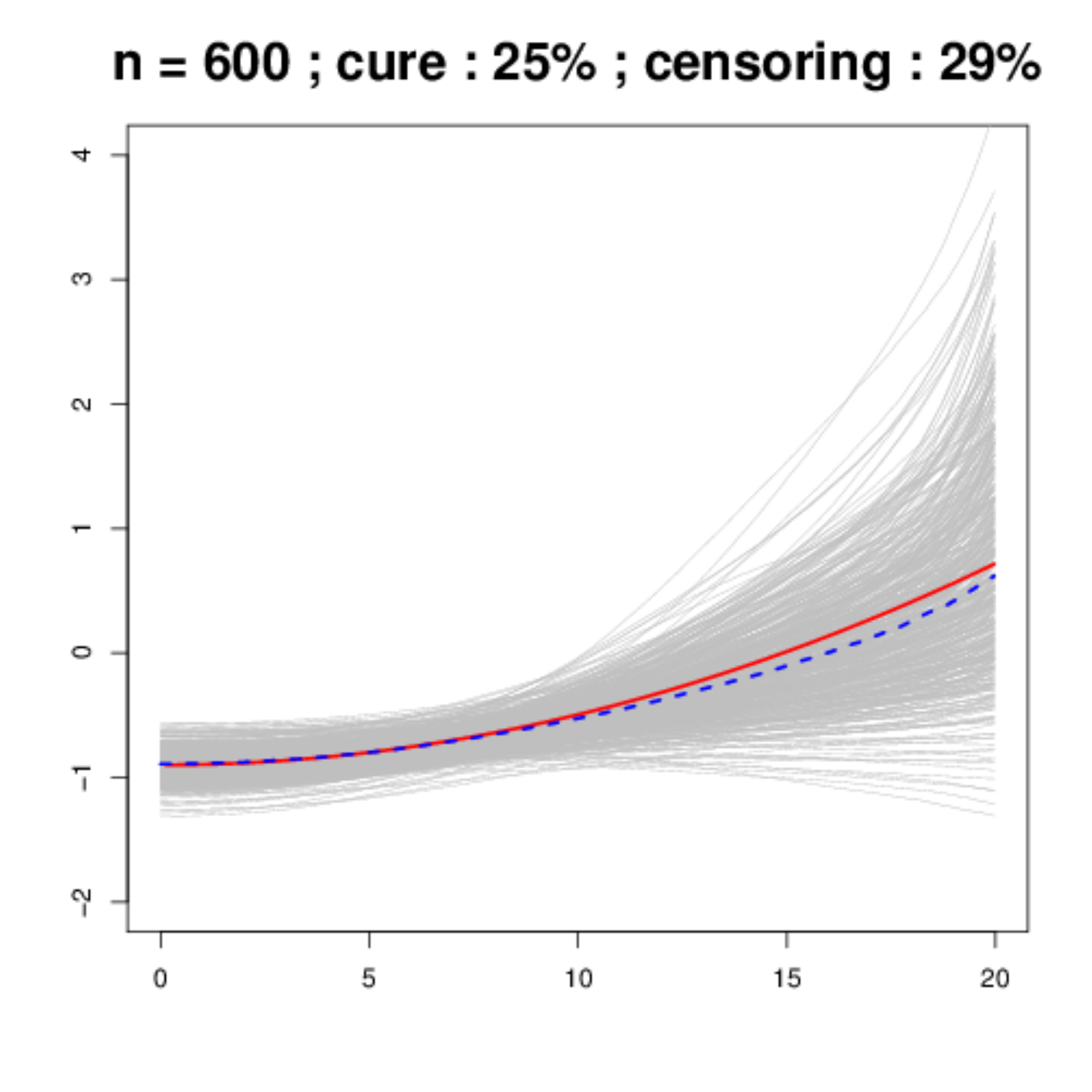}  \\
\includegraphics[width=5.35cm,height=5.35cm]{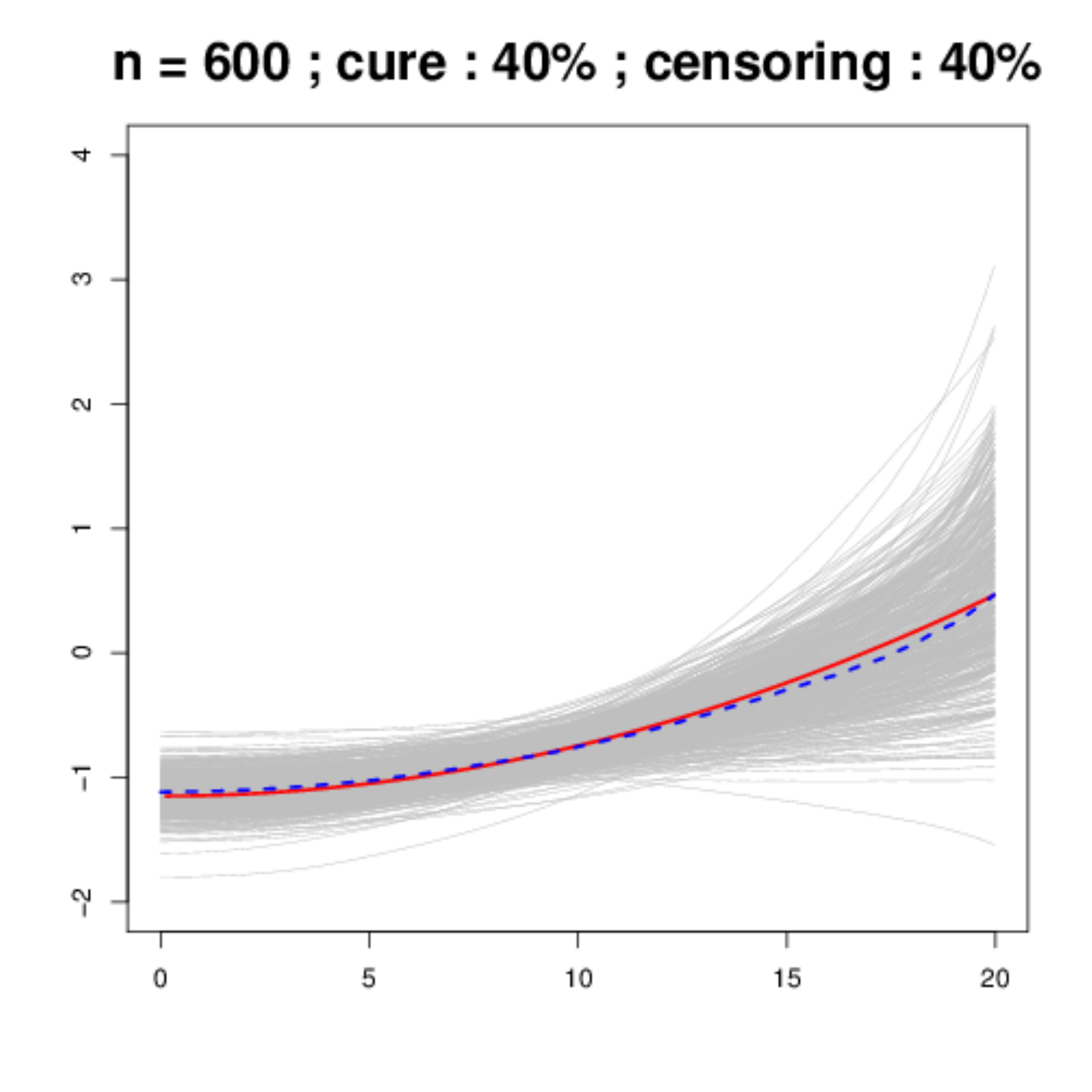} & 
\includegraphics[width=5.35cm,height=5.35cm]{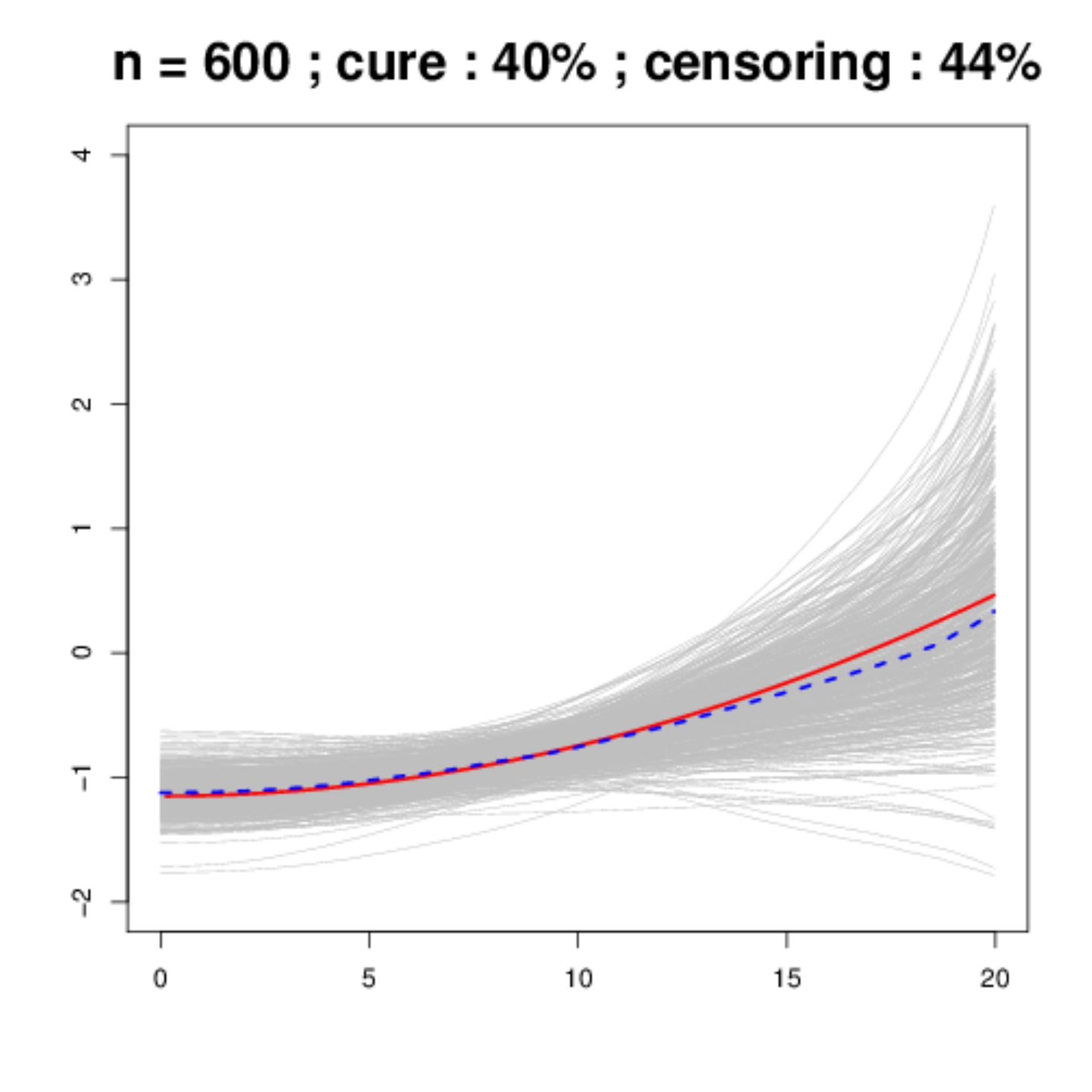}  
\end{tabular}
\caption{\footnotesize{Simulation results when the follow up is sufficiently long : estimation of the population log-hazard ratio $\log(HR_p(t))$ for $S=500$ replications (one grey curve per data set) and sample size $n=600$. Each row refers to a percentage of cured individuals (row 1 : 25\%, row 2 : 40\%) with different global right censoring rates (left : setting 1 ; right : setting 2). The solid line corresponds to the true function and the dashed line is the pointwize median of the 500 estimated curves. The hazard ratio is obtained by contrasting the groups induced by the binary covariate (for a median value of the continuous covariate.)}}
\label{Bremhorst::PopHazard600}
\end{center}
\end{figure}

\begin{figure}[H] 
\begin{center}
\begin{tabular}{cc}
\includegraphics[width=5.35cm,height=5.35cm]{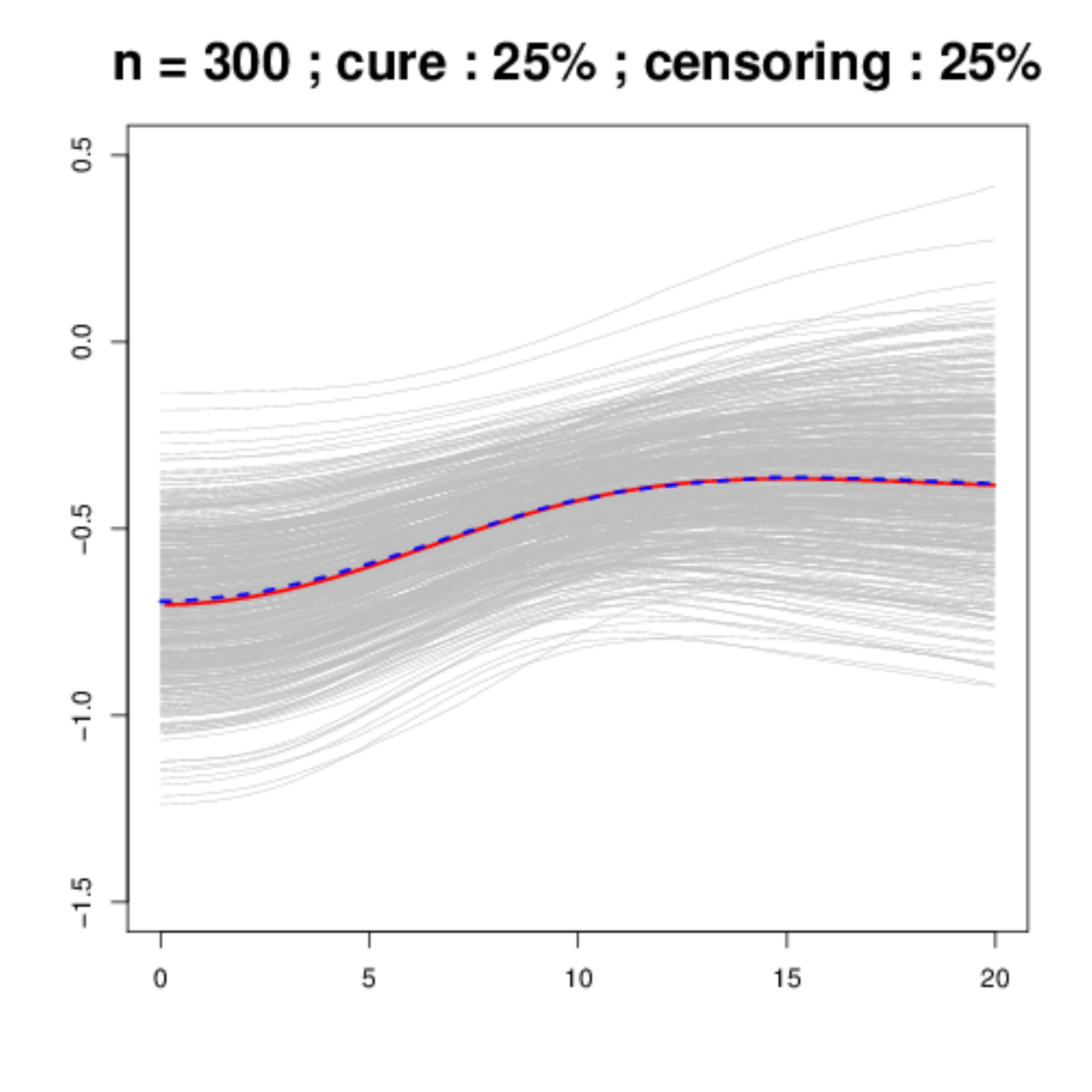} & \includegraphics[width=5.35cm,height=5.35cm]{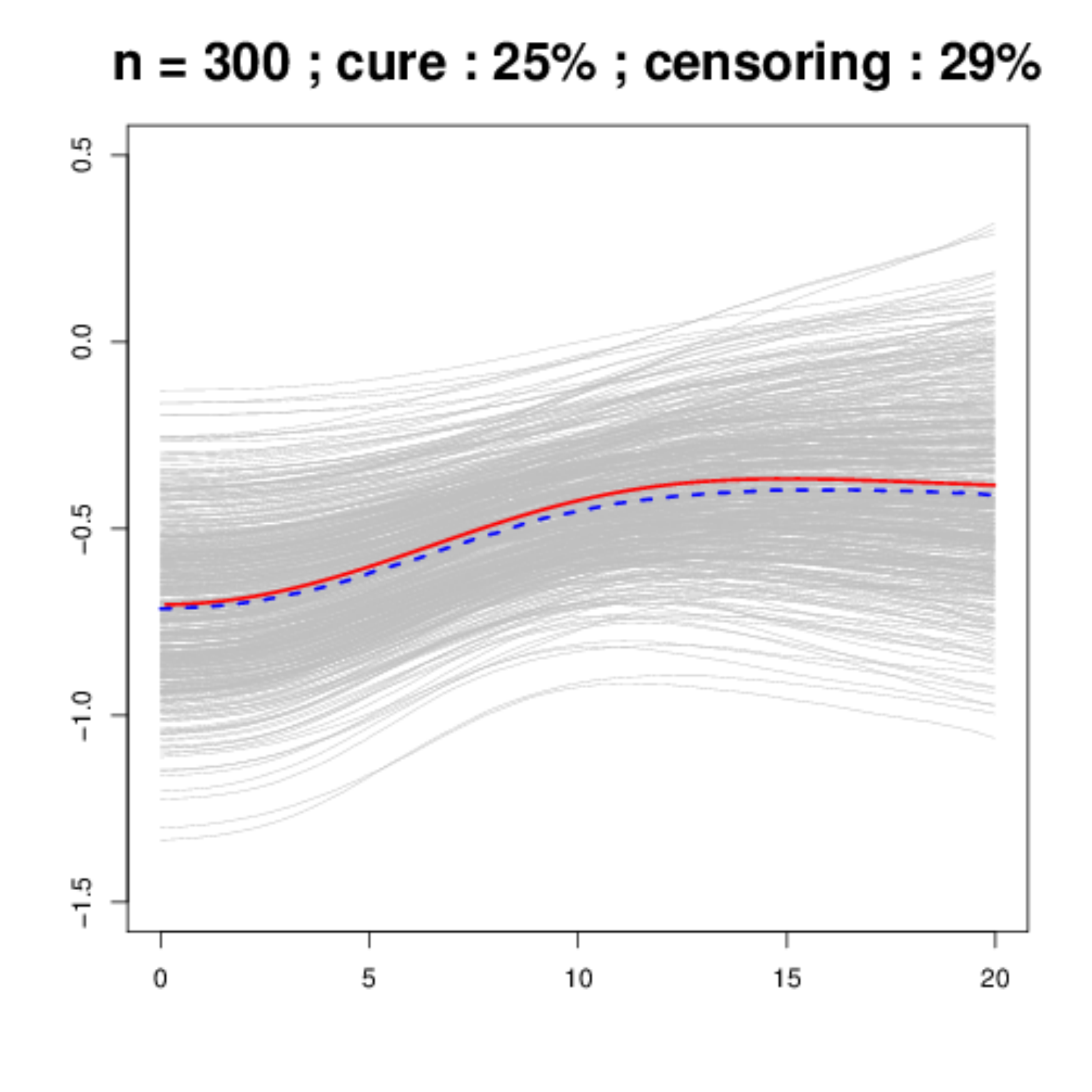}  \\
\includegraphics[width=5.35cm,height=5.35cm]{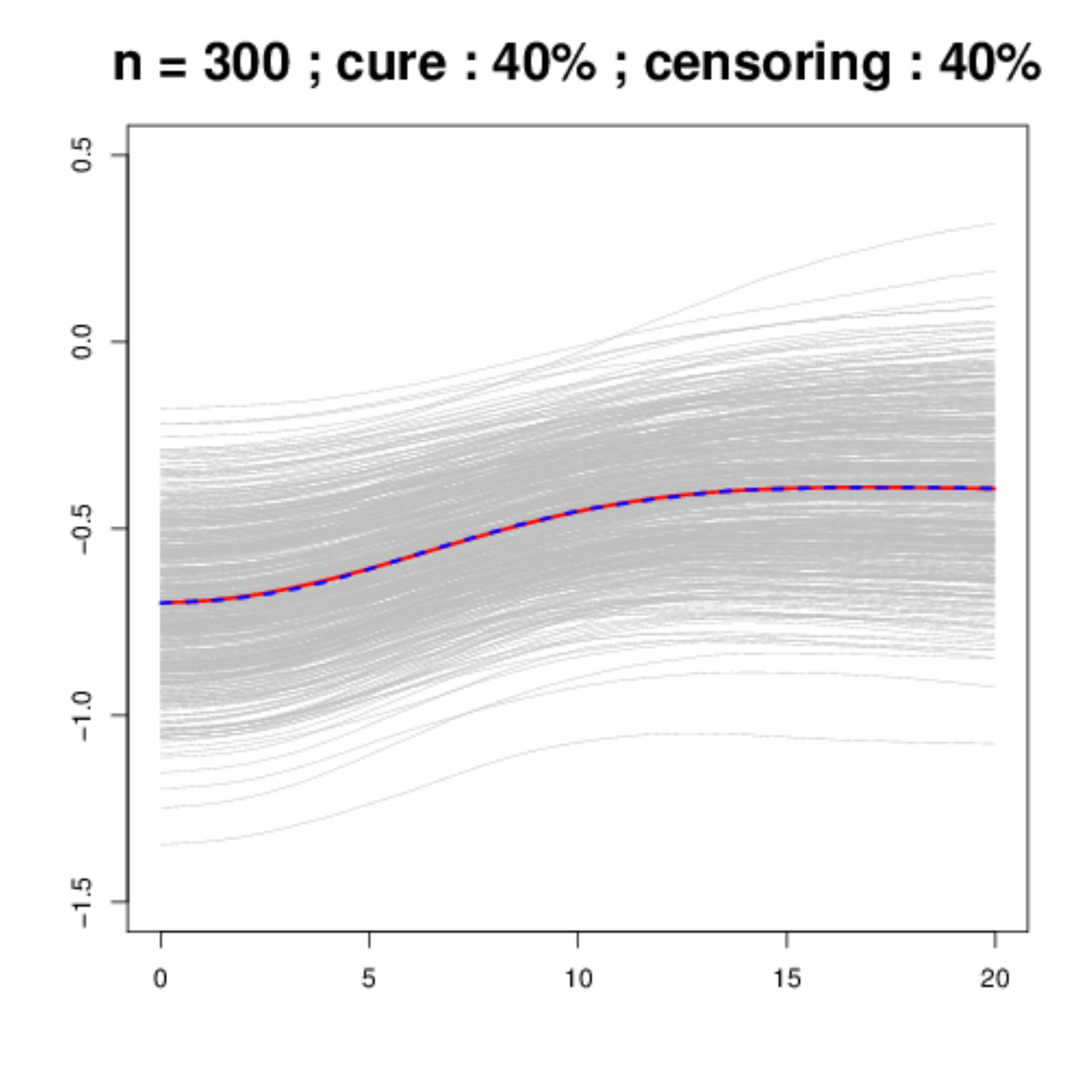} & \includegraphics[width=5.35cm,height=5.35cm]{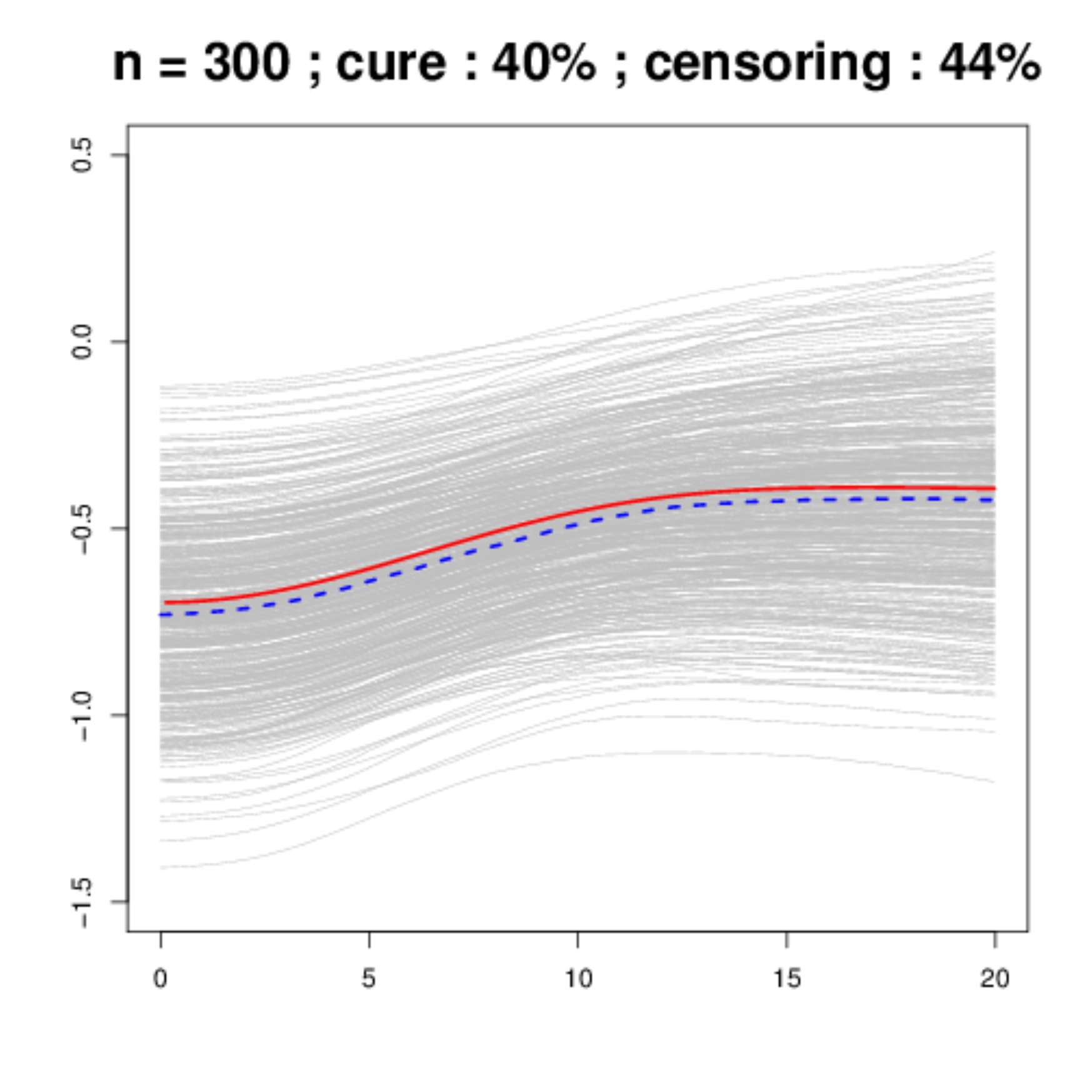}  
\end{tabular}
\caption{\footnotesize{Simulation results when the follow up is sufficiently long : estimation of the log-hazard ratio for the susceptible individuals $\log(HR_u(t))$ for $S=500$ replications (one grey curve per data set) and sample size $n=300$. Each row refers to a percentage of cured individuals (row 1 : 25\%, row 2 : 40\%) with different global right censoring rates (left : setting 1 ; right : setting 2). The solid line corresponds to the true function and the dashed line is the pointwize median of the 500 estimated curves. The hazard ratio is obtained by contrasting the groups induced by the binary covariate (for a median value of the continuous covariate.)}}
\label{Bremhorst::SuceptHazard300}
\end{center}
\end{figure}

\noindent
The accuracy of the proposed methodology was evaluated using simulations when the follow up is sufficiently long (see Section \ref{Bremhorst::Identifiability}). In each setting, the baseline distribution in (\ref{Bremhorst:modbothcov}) corresponds to a Weibull with mean $8$ and standard deviation $4.18$. Two covariates were included in the regression parts : $W_1 \sim N(0, 1)$ and $W_2 \sim Bernoulli(0.5)$. Since the sufficient follow up assumption is satisfied, both covariates can be used simultaneously to model the probability of being cured and the time necessary for a cell to yield a detectable tumor without causing an identifiability problem. Thus, we set $\mathbf{X} = \{W_1, W_2\}=\mathbf{Z}$. The regression coefficients associated to $W_1$ and $W_2$, in the Cox PH model, are set to 0.4 and -0.4, respectively. The upper bound for the observed failure time was set at $23$ as more than $99\%$ of the events occur before that time under the chosen Weibull distribution. Three percentages  were considered for the proportion of cured individuals : $15\%$, $25\%$ and $40\%$. The value of the regressors ($\beta_0, \mathbf{\beta}) $ were tuned to get these percentages. Each dataset was generated using the biological motivation of the model as follows : 
For each subject : 
\begin{enumerate}
\item[1) ] Generation of the number of carcinogenic cells using $N \sim Pois(\theta(\mathbf{x}))$ with $\theta(\mathbf{x})$ = $\exp(\beta_0 + \mathbf{x}^T \mathbf{\beta})$ ; 
\item[2) ] If $N \ne 0$, $N$ latent event times $Y_1, ..., Y_N$ are generated using the Cox proportional hazard model. The observed failure time is defined as $T= \min(Y_1, ..., Y_N)$. This step is repeated until $T < 23$. Note that it had to be repeated more than once in less than 1\% of the cases.
If $N=0$, the failure time for the cured individual is set to an arbitrary large value (999, say).  
\item[3) ] The global right censoring rate is controlled by one of the two following censoring distributions :  
\begin{enumerate}
\item[a) ] setting 1 : an uniform distribution on $[20, 25]$. This censoring distribution ensures that almost all the right censored subjects are cured and  identifiable (since their censoring time are located in the plateau of the Kaplan Meier estimate of the survival distribution).
\item[b) ] setting 2 : a Weibull distribution with mean $22.28$ and standard deviation $8.08$ truncated at $25$. Using this censoring distribution, the censoring time of only $25\%$ of the cured subjets are located in the plateau of the Kaplan Meier estimate and $4\%$ of the suceptible individuals are right censored. \\
\end{enumerate} 
\end{enumerate}
We use the model described in Section \ref{Bremhorst::FlexibleSpecification} with a cubic B-splines basis associated to 12 equidistant knots on $[0, t_{Rcens}]$, where $t_{Rcens}$ is equal to $25$ and a third order roughness penalty to counterbalance the flexibility of the B-splines. The simulations were performed on S = 500 replicates of sample size n = 300 and 600. \\
Using the procedure described in Section \ref{PosteriorMCMC}, we construct a chain of length $23000$ (including a burnin of $3000$) to explore the joint posterior distribution. The behavior and the convergence of the chains were assessed by an examination of the trace plots and using diagnostics tools such as in Geweke (1992).   \\
For the sake of brevity, we only report the results when the percentage of cured individuals is $25\%$ and $40\%$. Tables \ref{Bremhorst:ResultCov300} and \ref{Bremhorst:ResultCov600} summarize the simulation results for the regression parameters. One can see that the posterior medians (as estimators) of the regression coefficients show a negligible bias whatever the setting. The empirical standard error and RMSE of the posterior median of the regression parameters decrease slighly when the sample size increases and increase when the proportion of cured individuals with a censoring time greater than the maximum observed failure time decreases and when the percentage of right censoring in the non cured population increases. \\
In each setting, the coverage probabilities of the 90\% and 95\% credible intervals are close to their nominal value. The numerical results suggest that the proportion of cured individuals does not affect the accuracy of the estimates. \\
The estimates of the baseline survival function are plotted in Figures \ref{Bremhorst::baseline300} (when $n=300$) and \ref{Bremhorst::baseline600} (when $n=600$). The variability of the estimated baseline distribution increases slightly when the proportion of cured individuals with a censoring time greater than the maximum observed failure time decreases and when the percentage of right censoring in the non cured population increases and decreases when the sample size increases. A limited bias decreasing with sample size seems to appear in the estimation of the right tail of $S_0(t)$ when the proportion of cured individuals with a censoring time greater than the maximum observed failure time decreases and when the percentage of non cured right censored subjects increases. As for the regression parameters, the percentage of immune individuals does not affect the accuracy of the estimates. The same conclusions as for the regression parameters and for $S_0(t)$ hold for the population log-hazard ratio (see Figures \ref{Bremhorst::PopHazard300}, when $n=300$, and \ref{Bremhorst::PopHazard600}, when $n=600$), while the log-hazard ratio of the susceptible individuals is properly estimated whatever the setting (see Figure \ref{Bremhorst::SuceptHazard300}, when $n = 300$). These hazard ratios are obtained by contrasting the groups induced by the binary covariate (for a median value of the continuous covariate).
\begin{table}
\begin{center}
\caption{Simulation results for $S=500$ replicates and a sample size of $n=300$ when the follow up is sufficiently long. The percentage of cured individuals, the considered setting and the true value of the regression parameters are defined in the first three columns. The bias, the coverage of the 90\% and 95\% credible intervals, the empirical standard error (ESE) and the RMSE of the posterior median of the regression parameters are presented for each scenario.}
\label{Bremhorst:ResultCov300}
\begin{tabular}{cccccccc}
\hline
Cure & Setting & Parameters & Bias & $CV_{90\%}$ & $CV_{95\%}$ & ESE & RMSE \\ 
\hline
$\multirow{10}{*}{ 25\% }$& $\multirow{5}{*}{ 1 }$ & $\beta_0$ = 0.75 & 0.028 &88.4&93.8&0.130&0.018 \\
&& $\beta_1$ =  0.80& 0.016 & 90.8 &94.8&0.112&0.013 \\
&& $\beta_2$ = -0.50 & -0.013 & 91.6 &96.6&0.172&0.030 \\
&& $\gamma_1$ = 0.40& -0.035 & 91.0 &95.8 &0.138 &0.142 \\
&& $\gamma_2$ = -0.40 & 0.008 & 87.4  & 94.2 &0.221& 0.221 \\
\cline{3-8}
& $\multirow{5}{*}{ 2 }$ & $\beta_0$ = 0.75 & 0.052  &85.8&92.6&0.162&0.030 \\
&& $\beta_1$ =  0.80& -0.003&93.4 &97.8&0.129&0.017 \\
&& $\beta_2$ = -0.50 &0.017 & 88.6&94.0&0.209&0.044 \\
&& $\gamma_1$ = 0.40& 0.001 & 92.8 & 95.8& 0.171 & 0.171 \\
&& $\gamma_2$ = -0.40 & -0.029 & 88.2 & 93.6 &0.283&0.282  \\
\cline{3-8}
$\multirow{10}{*}{ 40\% }$& $\multirow{5}{*}{ 1 }$ & $\beta_0$ = 0.30 & 0.006 &90.6&94.0&0.128&0.017 \\
&& $\beta_1$= 1.00 & 0.013 & 90.6&95.0&0.119&0.014 \\
&& $\beta_2$ = -0.75 & -0.008 & 90.6 &94.4&0.182&0.033 \\
&& $\gamma_1$ = 0.40& -0.021 & 90.0 &95.6 & 0.150 & 0.151\\
&& $\gamma_2$ =  -0.40& 0.003 &92.0  & 96.8 &0.215& 0.215 \\
\cline{3-8}
& $\multirow{5}{*}{ 2 }$ & $\beta_0$ =  0.30& 0.043 &86.2&91.8&0.155&0.027 \\
&& $\beta_1$ = 1.00 & -0.014&93.6 &97.0&0.137&0.019 \\
&& $\beta_2$ = -0.75 & 0.024& 88.0&93.4&0.222&0.049 \\
&& $\gamma_1$ = 0.40& 0.005 & 91.8 & 96.4 & 0.183 & 0.183 \\
&& $\gamma_2$ =  -0.40& -0.025 & 90.4 & 94.4 &0.276& 0.276 \\
\hline
\end{tabular}
\end{center}
\end{table}

\begin{table}
\begin{center}
\caption{Simulation results for $S=500$ replicates and a sample size of $n=600$ when the follow up is sufficiently long. The percentage of cured individuals, the considered setting and the true value of the regression parameters are defined in the first three columns. The bias, the coverage of the 90\% and 95\% credible intervals, the empirical standard error (ESE) and the RMSE of the posterior median of the regression parameters are presented for each scenario.}
\label{Bremhorst:ResultCov600}
\begin{tabular}{cccccccc}
\hline
Cure & Setting & Parameters & Bias & $CV_{90\%}$ & $CV_{95\%}$ & ESE & RMSE \\ 
\hline
$\multirow{10}{*}{ 25\% }$& $\multirow{5}{*}{ 1 }$ & $\beta_0$ = 0.75 & 0.020 &90.2&94.0&0.087&0.008 \\
&& $\beta_1$ =  0.80&0.017 &90.4 &95.2&0.077&0.006 \\
&& $\beta_2$ = -0.50 &-0.020 & 91.6&95.8&0.115&0.014 \\
&& $\gamma_1$ = 0.40& -0.038 & 87.6 & 93.6 & 0.097 & 0.106 \\
&& $\gamma_2$ = -0.40 & 0.022 & 89.8 & 96.4 &0.139& 0.141  \\
\cline{3-8}
& $\multirow{5}{*}{ 2 }$ & $\beta_0$ = 0.75 & 0.036 &87.2&92.8&0.107&0.013 \\
&& $\beta_1$ =  0.80&0.009 & 93.2 &97.2&0.090& 0.008 \\
&& $\beta_2$ = -0.50 & -0.011 & 88.6 &93.6&0.150& 0.023 \\
&& $\gamma_1$ = 0.40& -0.023 & 91.0  & 96.0 & 0.115 &  0.118\\
&& $\gamma_2$ = -0.40 & 0.018 & 91.6 & 95.0 &0.182& 0.183 \\
\cline{3-8}
$\multirow{10}{*}{ 40\% }$& $\multirow{5}{*}{ 1 }$ & $\beta_0$ = 0.30 & 0.005  &91.2&95.2&0.087&0.008 \\
&& $\beta_1$ = 1.00 &0.016 &91.2&95.6&0.079&0.006 \\
&& $\beta_2$ = -0.75 &-0.003 &91.6 &95.6&0.119&0.014 \\
&& $\gamma_1$ = 0.40& -0.027 & 91.2 &95.0 & 0.098 & 0.101\\
&& $\gamma_2$ =  -0.40& 0.036& 92.4 & 95.2 & 0.142&0.146  \\
\cline{3-8}
& $\multirow{5}{*}{ 2 }$ & $\beta_0$ =  0.30& 0.021  &89.4&94.8&0.094&0.009 \\
&& $\beta_1$ = 1.00 & -0.004 & 92.8 &95.4&0.093&0.009 \\
&& $\beta_2$ = -0.75 & 0.010& 91.2 &95.8 &0.142&0.020 \\
&& $\gamma_1$ = 0.40& -0.002 & 93.6 & 97.8 & 0.115 &0.115 \\
&& $\gamma_2$ =  -0.40& 0.025 & 92.6 & 96.4 &0.184&0.184  \\
\hline
\end{tabular}
\end{center}
\end{table}


\newpage
\subsection{Not sufficiently long follow up}
\begin{figure}[H] 
\begin{center}
\begin{tabular}{cc}
\includegraphics[width=5.35cm,height=5.35cm]{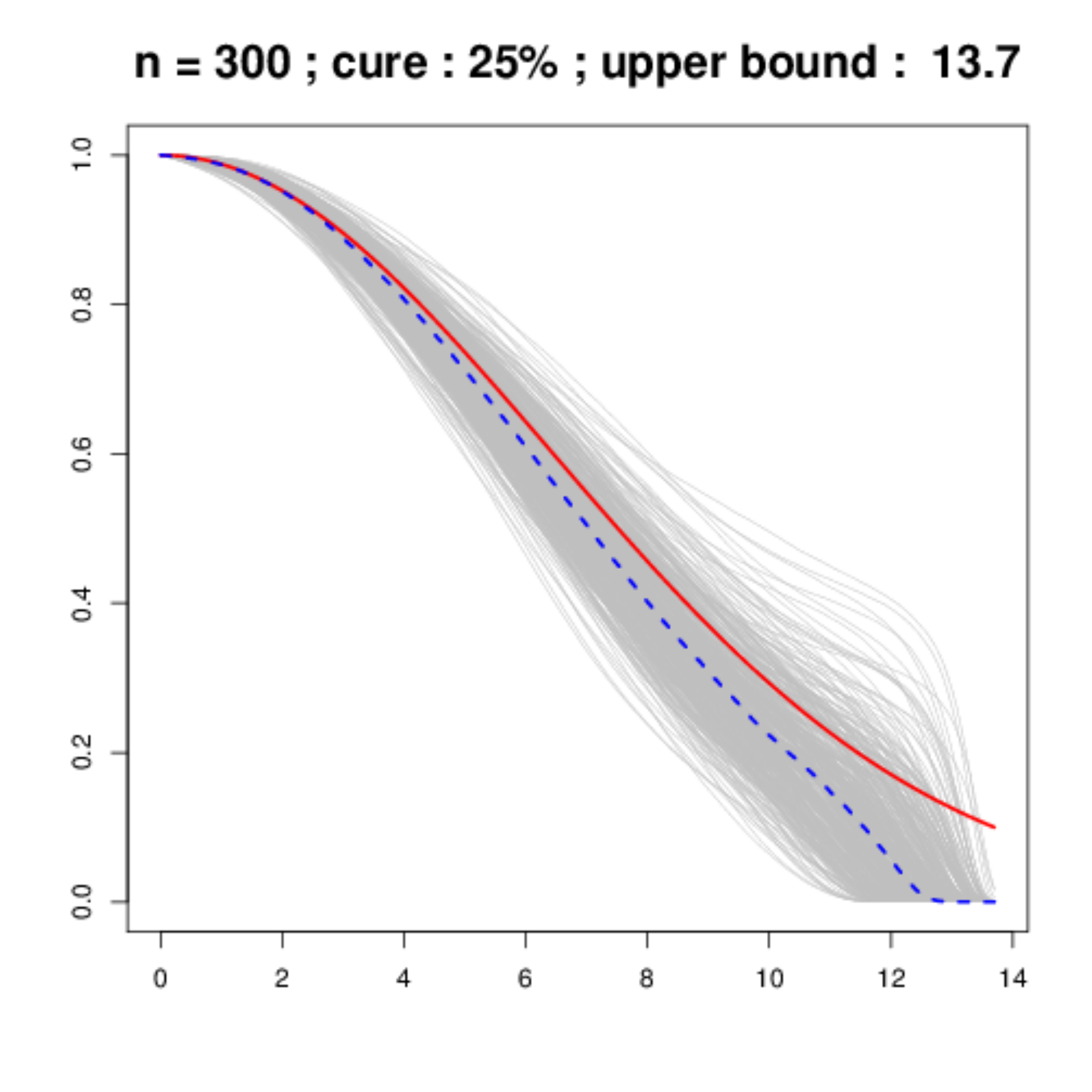} & 
\includegraphics[width=5.35cm,height=5.35cm]{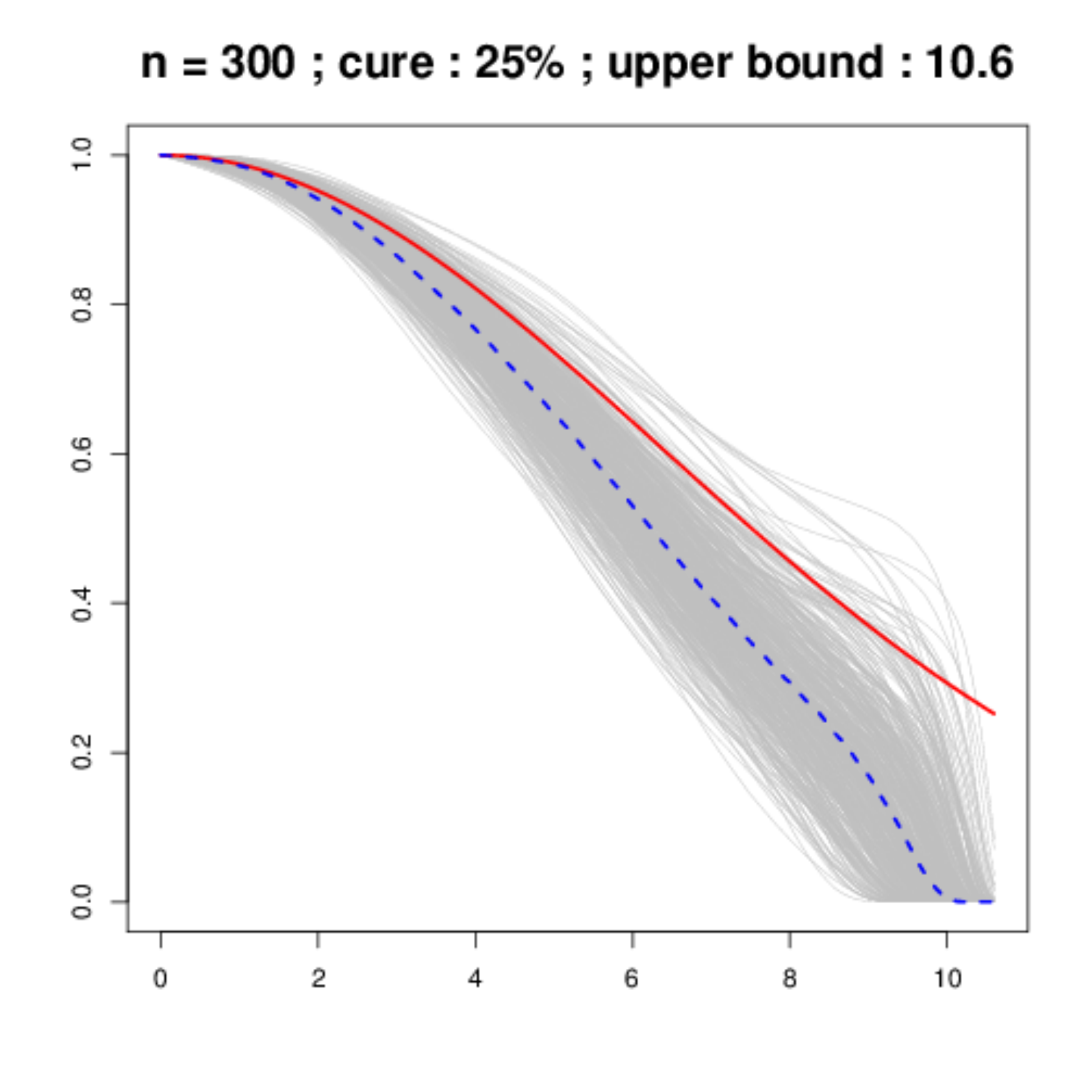}  \\
\includegraphics[width=5.35cm,height=5.35cm]{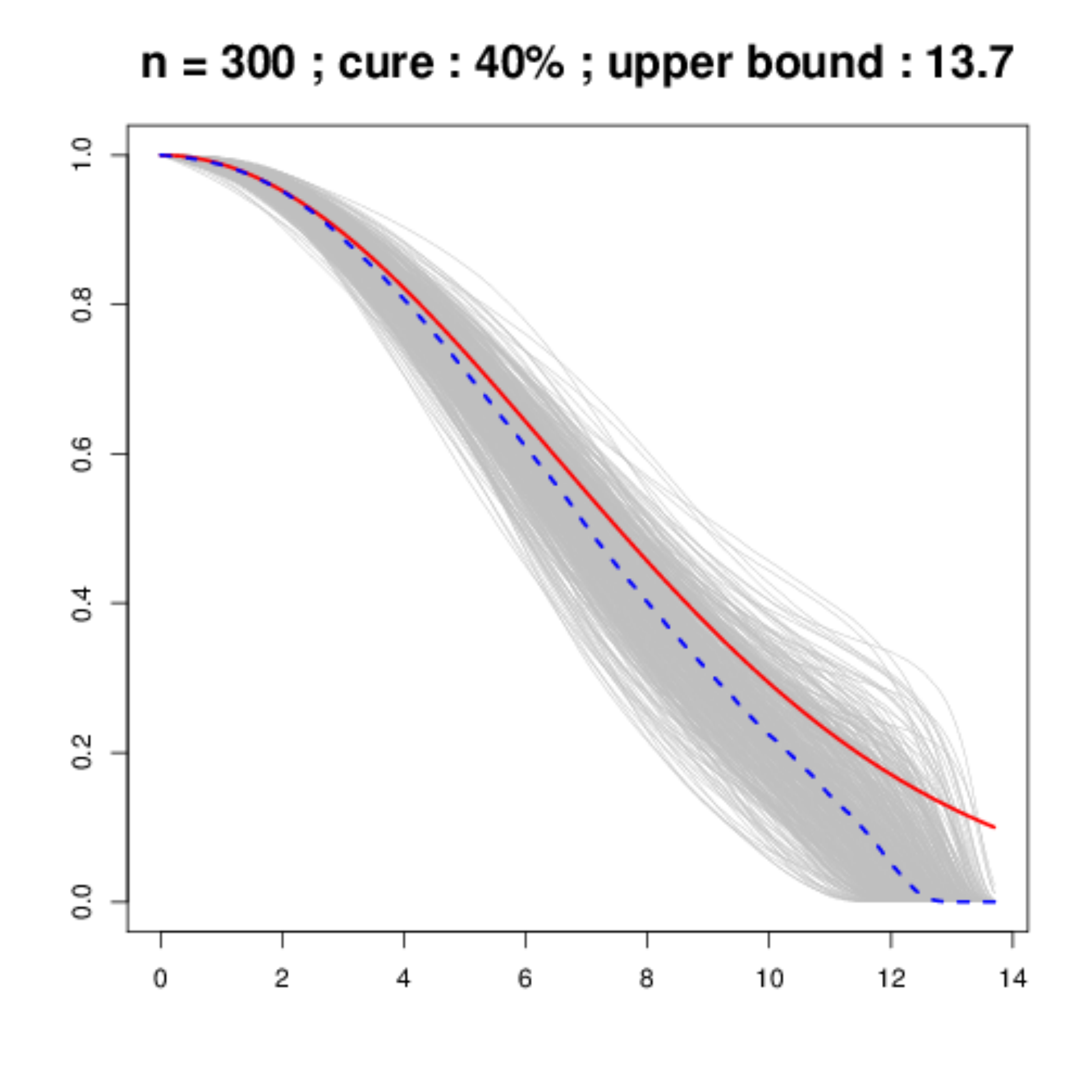} & 
\includegraphics[width=5.35cm,height=5.35cm]{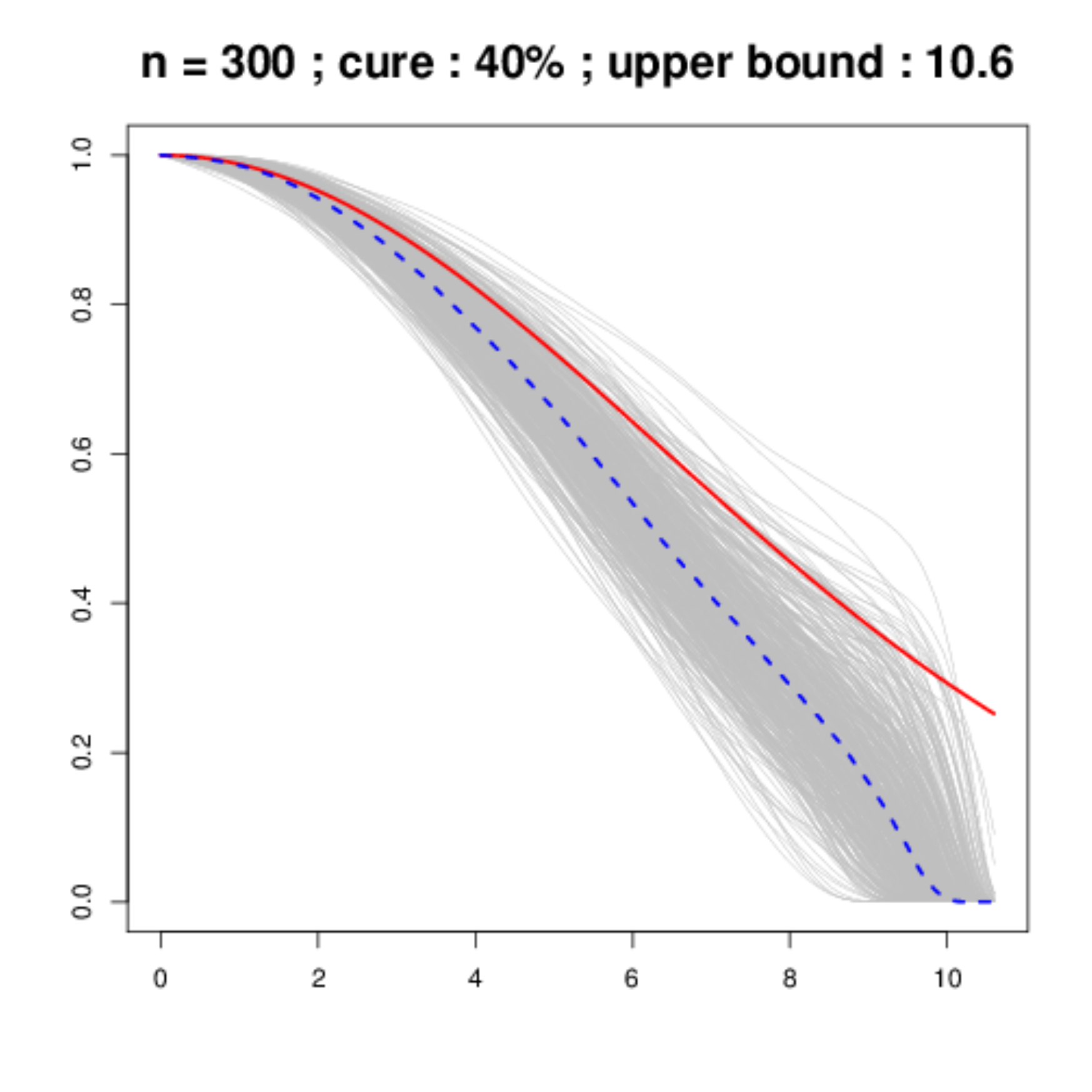}  \\
\end{tabular}
\caption{\footnotesize{Simulation results when the follow up is not sufficiently long : estimation of the baseline distribution $S_0(t)$ for $S=500$ replications (one grey curve per data set) and sample size $n=300$. Each row refers to a percentage of cured individuals (row 1 : 25\%, row 2 : 40\%) and the columns refer to the upper bound of the follow up (col 1 (setting 3) : $13.7$, col 2 (setting 4) : $10.6$). The solid line corresponds to the true function and the dashed line is the pointwize median of the 500 estimated curves.}}
\label{Bremhorst::NSLbaseline300}
\end{center}
\end{figure}

\begin{figure}[H] 
\begin{center}
\begin{tabular}{cc}
\includegraphics[width=5.35cm,height=5.35cm]{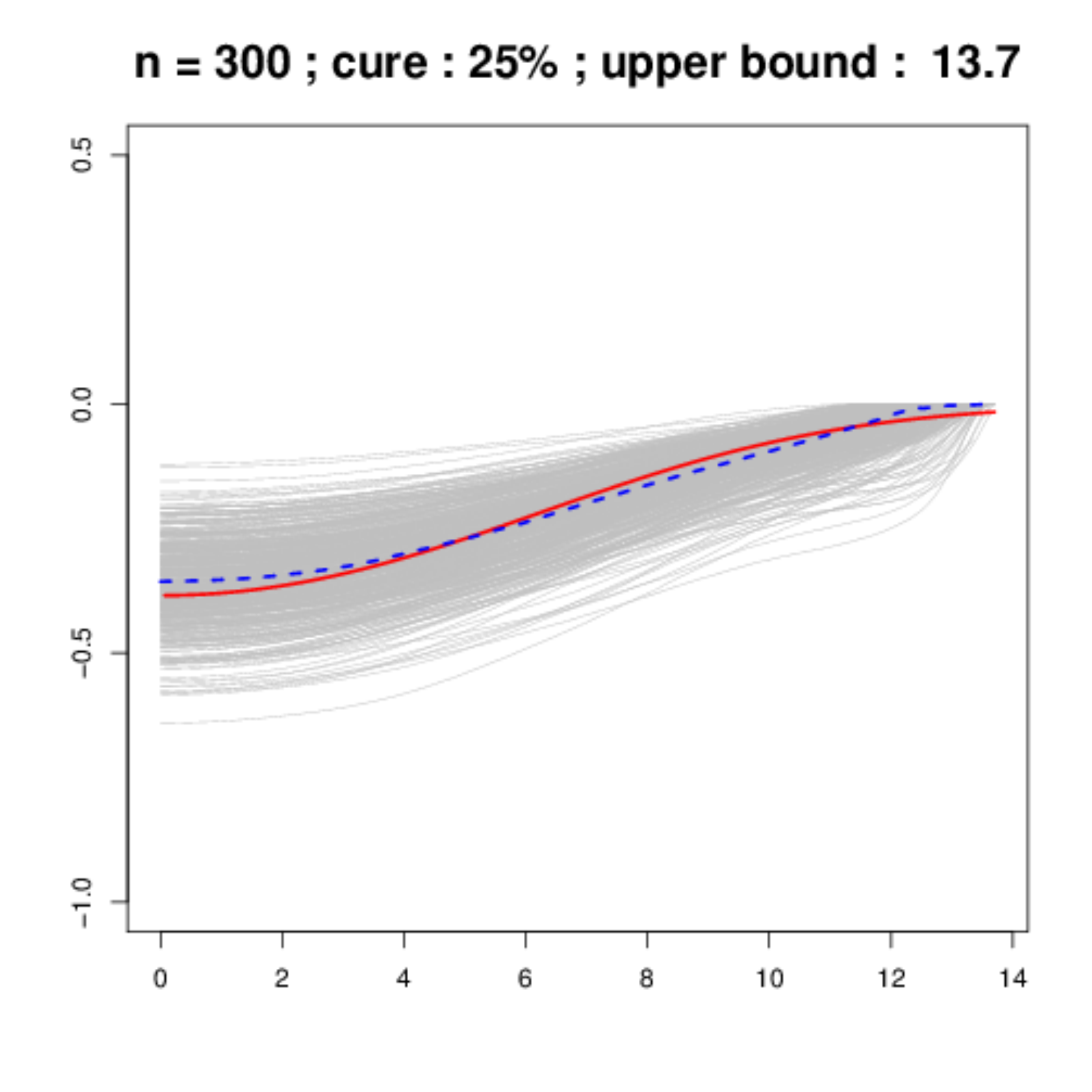} & \includegraphics[width=5.35cm,height=5.35cm]{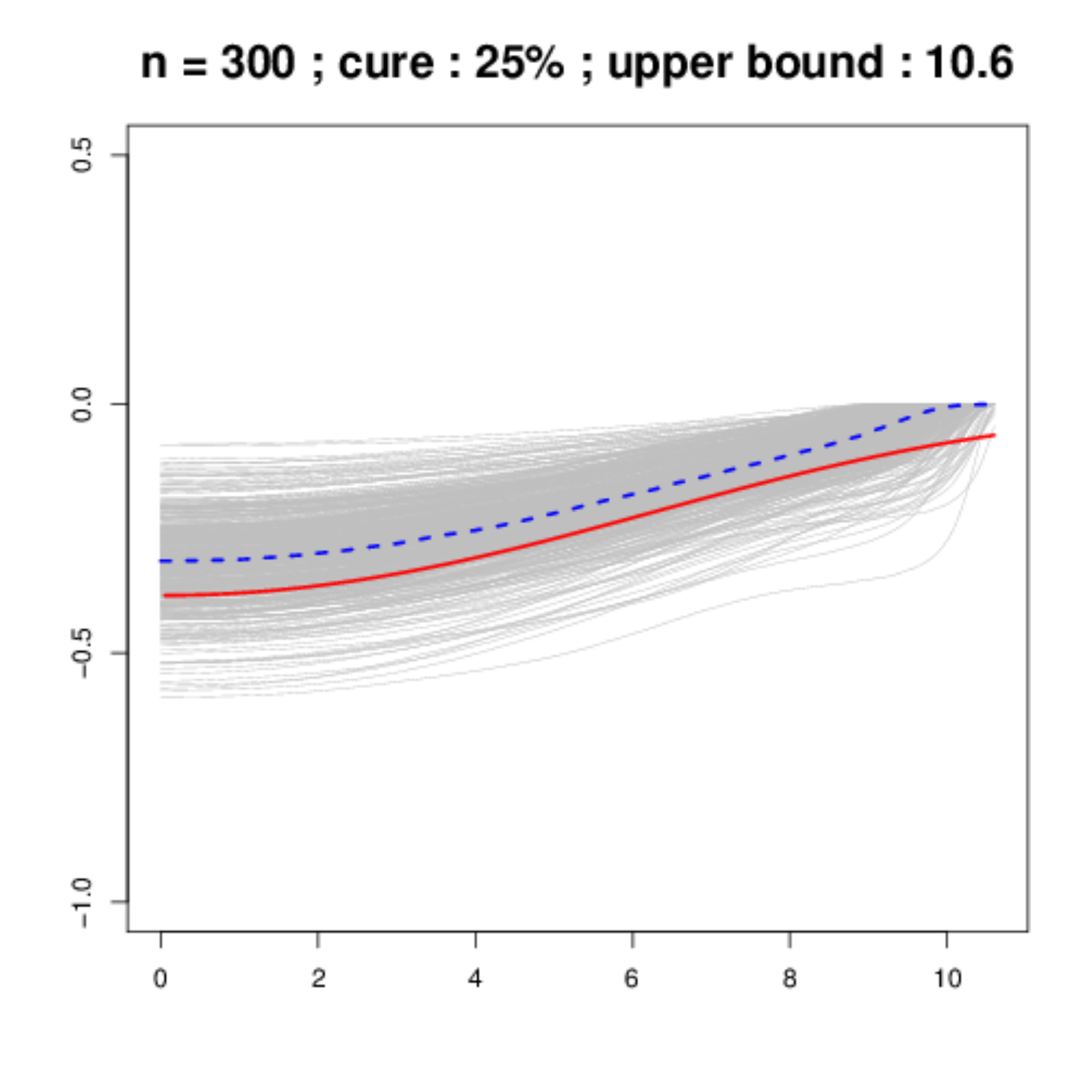}  \\
\includegraphics[width=5.35cm,height=5.35cm]{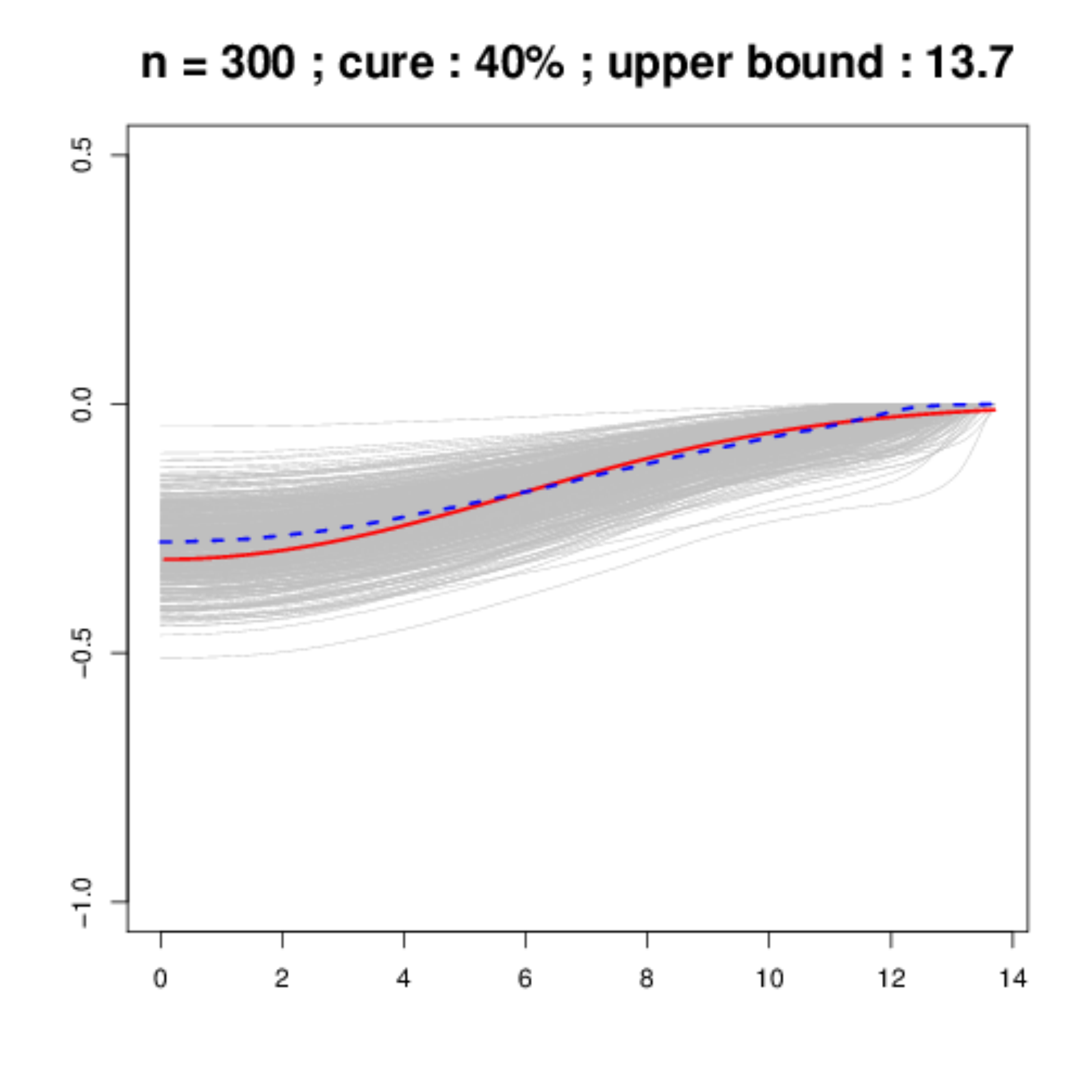} & \includegraphics[width=5.35cm,height=5.35cm]{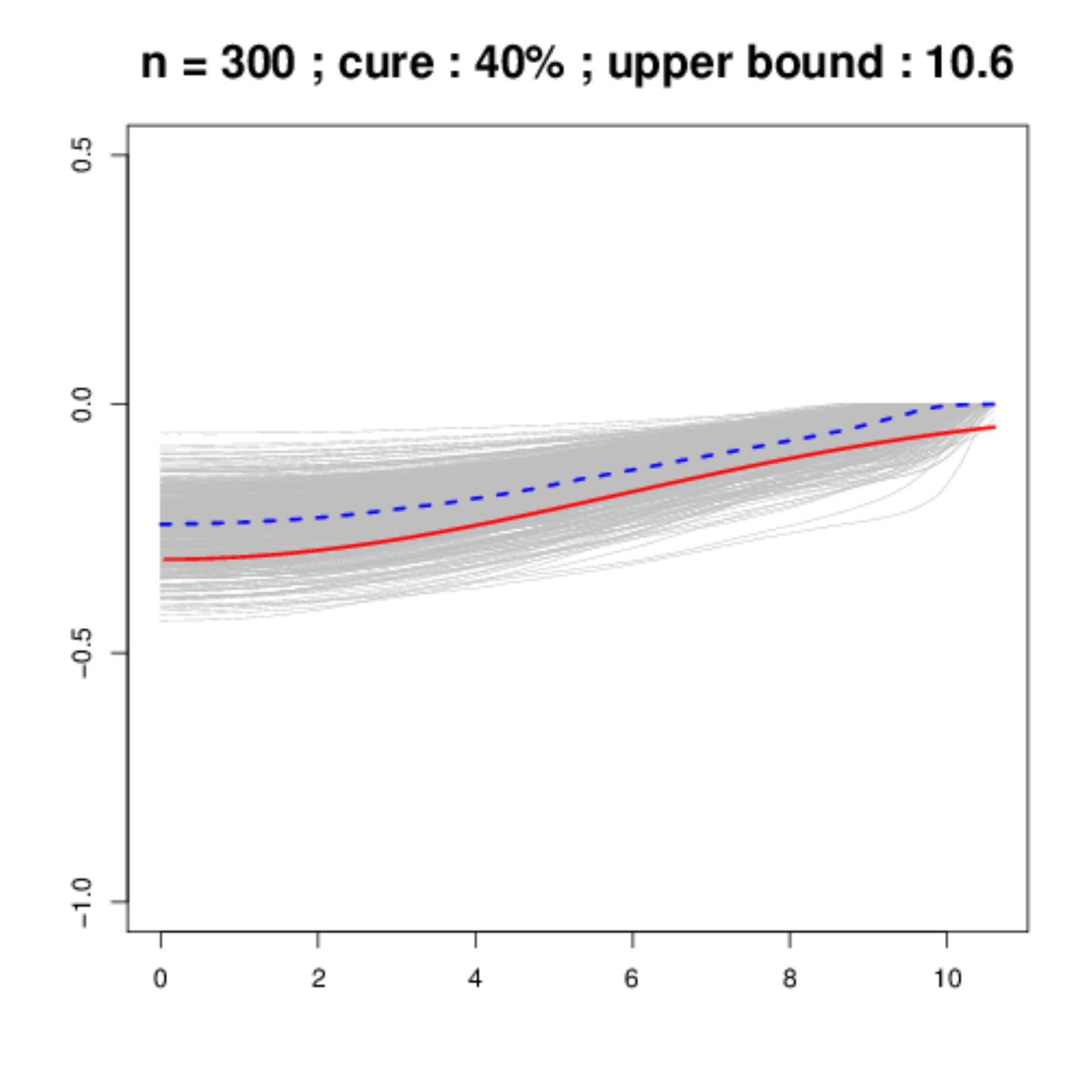}  
\end{tabular}
\caption{\footnotesize{Simulation results when the follow up is not sufficiently long : estimation of the log-hazard ratio for the susceptible individuals $\log(HR_u(t))$ for $S=500$ replications (one grey curve per data set) and sample size $n=300$. Each row refers to a percentage of cured individuals (row 1 : 25\%, row 2 : 40\%) and the columns refer to the upper bound of the follow up (col 1 (setting 3) : $13.7$, col 2 (setting 4) : $10.6$). The solid line corresponds to the true function and the dashed line is the pointwize median of the 500 estimated curves. The hazard ratio is obtained by contrasting the groups induced by the binary covariate (for a median value of the continuous covariate.)}}
\label{Bremhorst::NSLSuceptHazard300}
\end{center}
\end{figure}

\noindent
The restricted use of the promotion time model when the follow up of the study is not sufficiently long is illustrated using simulations. The datasets are generated using the procedure described in Section \ref{Bremhorst::Sufficientlylongfollowup}. However, since the follow up of the study is not sufficiently long, for identification purpose, $W_1$ only influences the time for a cell to yield a detectable tumor ($Z = W_1$), and $W_2$ only influences the probability of being cured ($X = W_2$). Note that if the hazard ratio is obtained by contrasting the groups induced by the binary covariate (for a median value of the continuous covariate), the population hazard ratio has, in this setting, a proportional hazard structure.  Three percentages were considered for the proportion of cure individuals : $15\%$, $25\%$ and $40\%$. To ensure that the largest censoring time is smaller than the largest observed failure time (which is the feature of non sufficiently long follow up), a Weibull distribution with mean $17.9$ and standard deviation $6.5$ truncated at $13.7$ (setting 3) and at $10.6$ (setting 4) is considered for the censoring distribution. \\
As in Section \ref{Bremhorst::Sufficientlylongfollowup}, we use the model described in Section \ref{Bremhorst::FlexibleSpecification} with a cubic B-splines basis associated to $12$ equidistant knots on $[0, t_{Rcens}]$, where $t_{Rcens}$ is equal to $13.7$ or $10.6$ depending on the considered censoring distribution and a third order roughness penalty to counterbalance the flexibility of the B-splines. The simulations were performed on $S = 500$ replicates of sample size $n = 300$ and $600$. A chain of length $23000$ (including a burnin of $3000$) is constructed using the procedure described in Section \ref{PosteriorMCMC}. As previously, traces and z-scores of all the model parameters are examinated to check the convergence of the MCMC algorithm. For the sake of brevity, we only report the results when the percentage of cured individuals is $25\%$ and $40\%$ and when the sample size is $300$. Table \ref{Bremhorst:NSLResultCov300} summarizes the simulation results for the regression parameters. As expected from the theory, in each setting, the posterior medians, as estimators of the intercept, show an underestimation. The biases are close to $\log(F_0(t_{Rcens}))$, where $F_0(.)$ is the cumulative distribution function of the considered baseline Weibull distribution in (\ref{Bremhorst:modbothcov}),  as can be explained from equation (\ref{Bremhorst::NSLintercept}). The posterior medians of $\beta_1$ and $\gamma_1$ show a non significant bias whatever the setting. The accuracy of the estimators of the regression parameters increases with the upper bound of the follow up and with sample size. In each setting, the coverage probabilities of the $90\%$ and $95\%$ credible intervals are close to their nominal value except for the intercept due to its underestimation.
Figure \ref{Bremhorst::NSLbaseline300} shows that the baseline distribution $S_0(t)$ is underestimated : this is due to the zero tail constraint. As illustrated on Figure \ref{Bremhorst::NSLSuceptHazard300}, an overestimation appears in the estimation of the log-hazard ratio of the suceptible individuals when the upper bound of the follow up is really to small. It happens when the baseline distribution function in (\ref{Bremhorst:modbothcov}) at the maximum possible censoring time is much smaller than $1$ : one has $F_0(10.6) = 0.75$ for the shortest follow up (setting 4) and $F_0(13.7) = 0.9$ in the most favorable setting (setting 3). Similar conclusions can be drawn when the sample size is equal to $600$. These simulation results corroborate the theoretical results proved in Lemma 1.

\begin{table}
\begin{center}
\caption{Simulation results for $S=500$ replicates and a sample size of $n=300$ when the follow up is not sufficiently long. The percentage of cured individuals, the considered setting and the true value of the regression parameters are defined in the first three columns. The bias, the coverage of the 90\% and 95\% credible intervals, the empirical standard error (ESE) and the RMSE of the posterior median of the regression parameters are presented for each scenario.}
\label{Bremhorst:NSLResultCov300}
\begin{tabular}{cccccccc}
\hline
Cure & Setting & Parameters & Bias & $CV_{90\%}$ & $CV_{95\%}$ & ESE & RMSE \\ 
\hline
$\multirow{6}{*}{ 25\% }$& $\multirow{3}{*}{ 3 }$ & $\beta_0$ = 0.70 & -0.093 &80.6&87.8&0.129& 0.025 \\
&& $\beta_1$ =  -0.70& 0.007 &88.4  &95.6&0.156&0.024 \\
&& $\gamma_1$ = 0.40& -0.029 & 86.6 &92.2& 0.111&0.114 \\
\cline{3-8}
& $\multirow{3}{*}{ 4 }$ & $\beta_0$ = 0.70 & -0.266 &46.6&60.6&0.157&0.096 \\
&& $\beta_1$ =  -0.70& 0.007 &91.0&93.8&0.169&0.029 \\
&& $\gamma_1$ = 0.40& -0.041 & 85.6&91.0& 0.124  & 0.130\\
\cline{3-8}
$\multirow{6}{*}{ 40\% }$& $\multirow{3}{*}{ 3 }$ & $\beta_0$ = 0.30 & -0.112 &77.4&85.0&0.142&0.033 \\
&& $\beta_1$= -0.80 &0.024 &91.0&96.0&0.178&0.032 \\
&& $\gamma_1$ = 0.40& -0.039 & 86.8 &93.0& 0.127 & 0.133 \\
\cline{3-8}
& $\multirow{3}{*}{ 4 }$ & $\beta_0$ =  0.30&-0.292  &45.4&56.6&0.159&0.104 \\
&& $\beta_1$ = -0.80 & 0.024&92.0&96.2&0.195&0.038 \\
&& $\gamma_1$ = 0.40& -0.049 & 86.2& 92.2 & 0.137 &0.145 \\
\cline{3-8}
\end{tabular}
\end{center}
\end{table}

\section{Application} 

\begin{figure}[H]
\begin{center}
\begin{tabular}{cc}
\includegraphics[width=5.35cm,height=5.35cm]{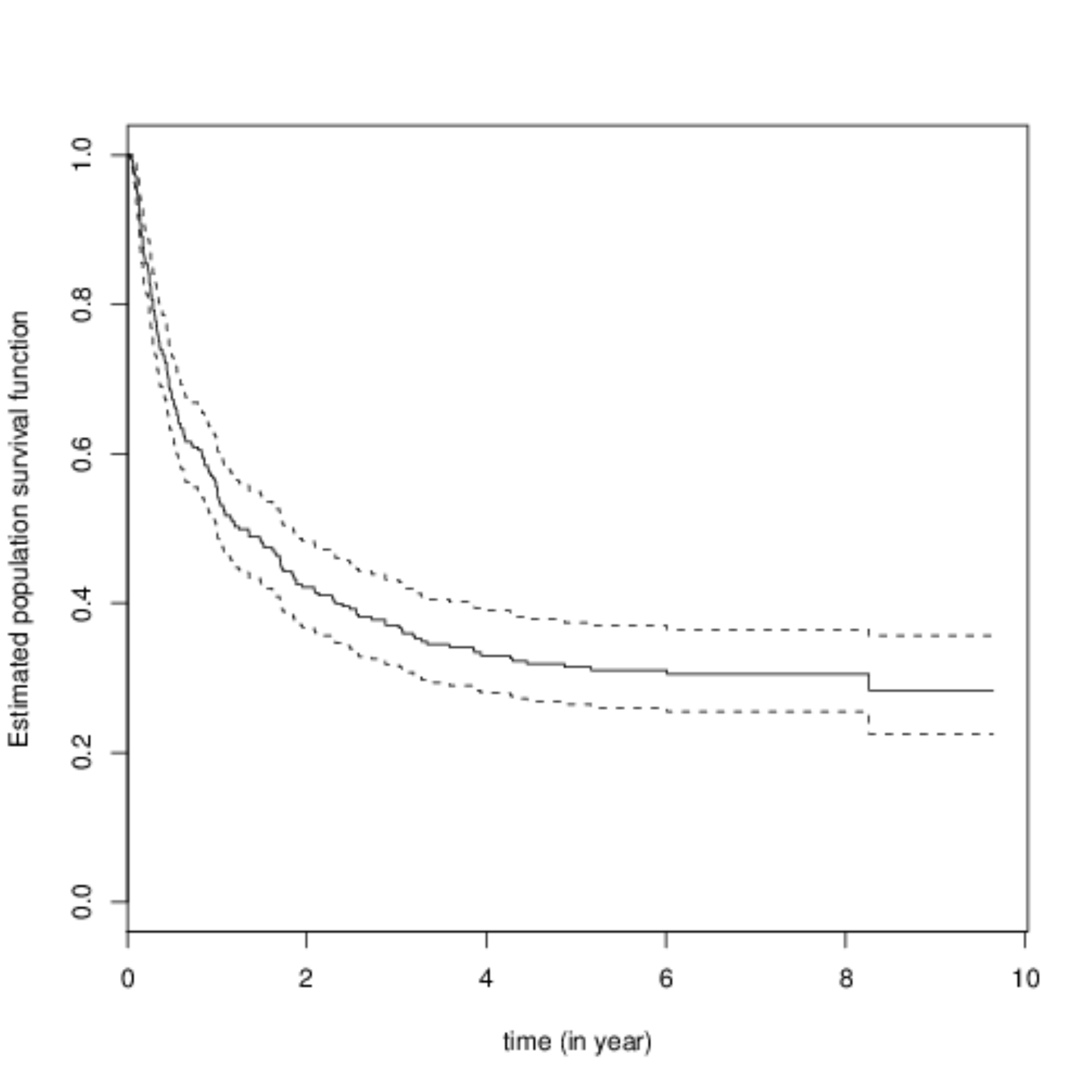}&
\includegraphics[width=5.35cm,height=5.35cm]{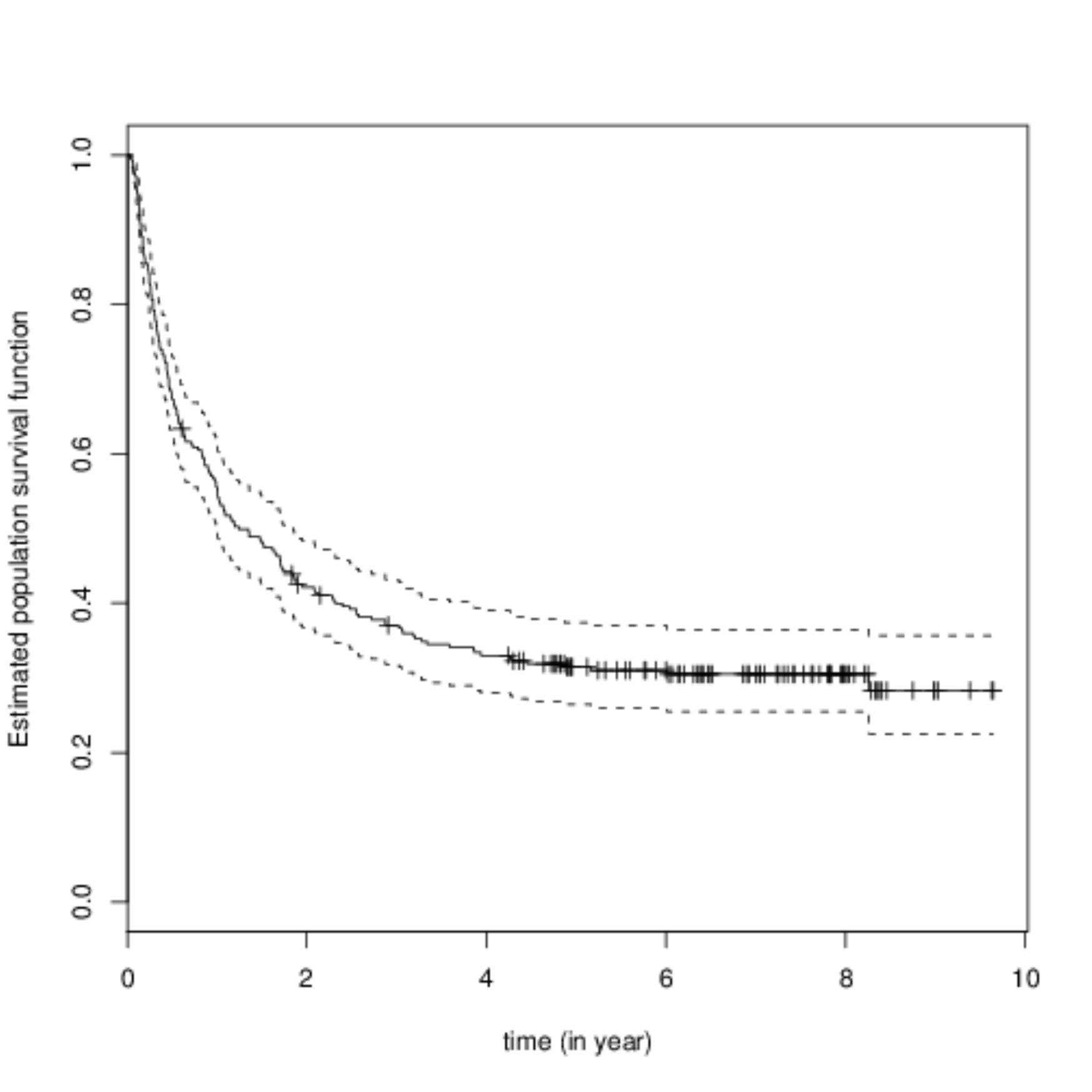}\\
\end{tabular}
\end{center}
\caption{Kaplan Meier estimated curve. Right : Each censored individual is marked by a cross. Left : Without cross for the censored individuals.}
\label{Bremhorst:KMpopulation}
\end{figure}
\begin{figure}[H] 
\begin{center}
\begin{tabular}{cc}
{
\includegraphics[width=5.35cm,height=5.35cm]{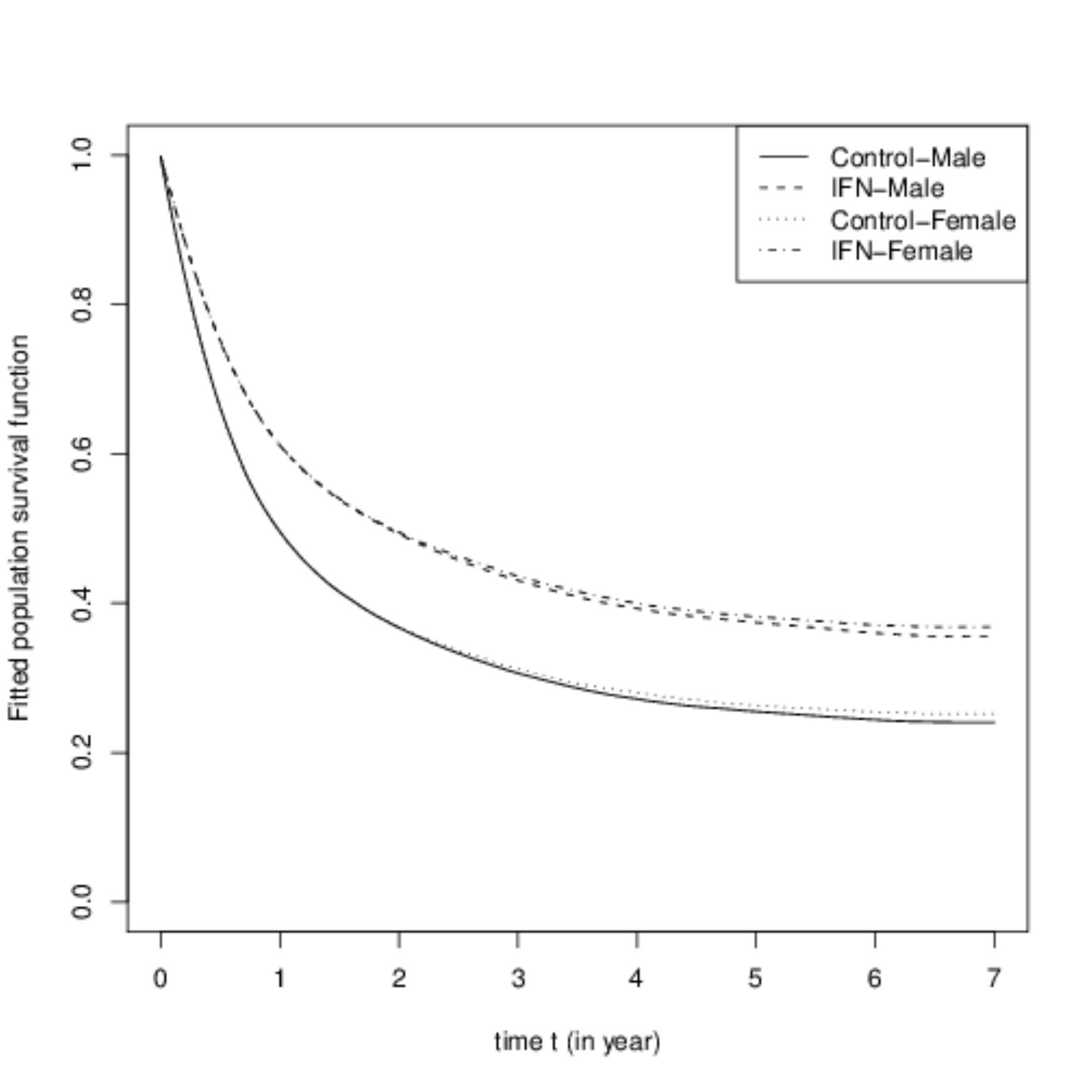} 
}
&
{
\includegraphics[width=5.35cm,height=5.35cm]{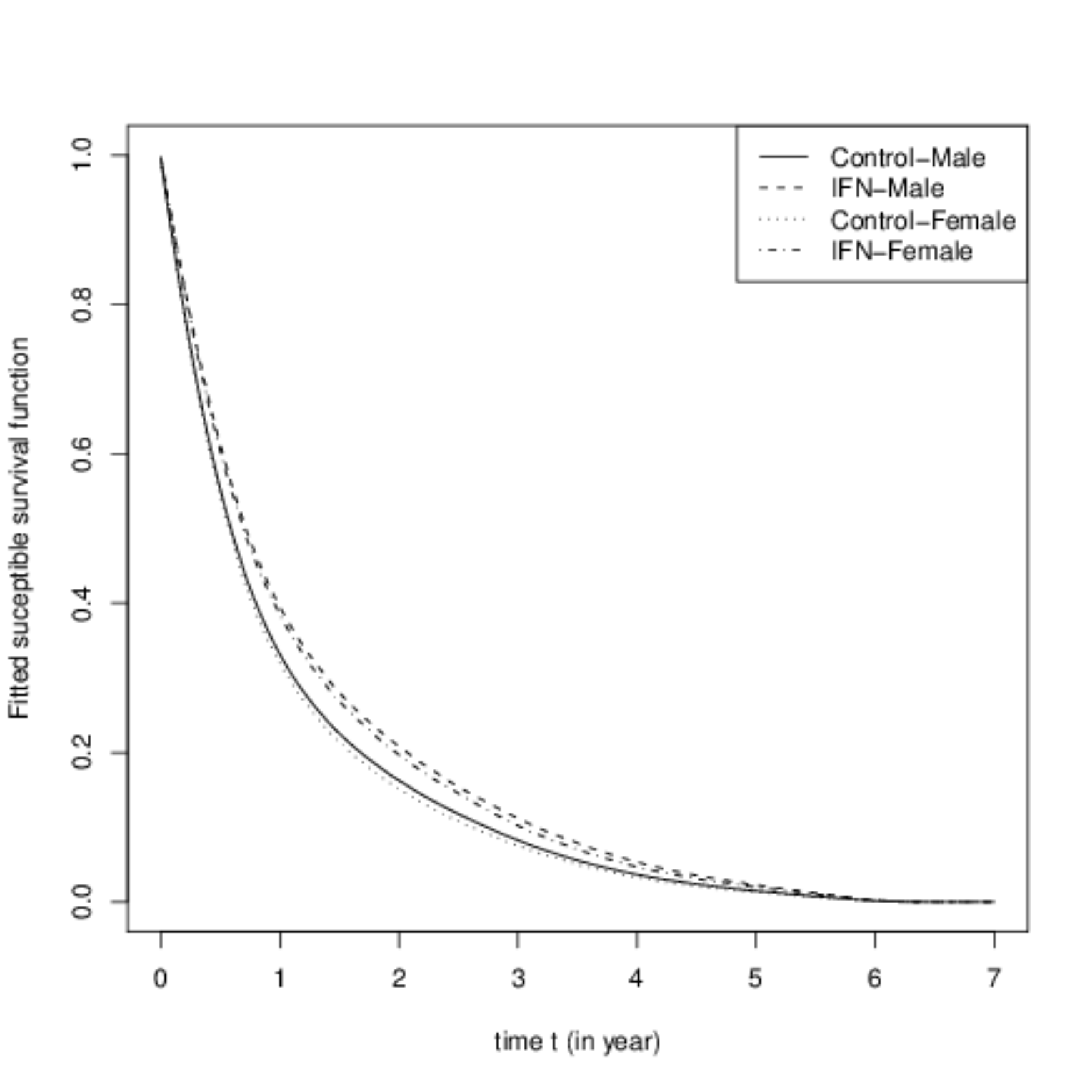} 
}
\end{tabular}
\caption{\footnotesize{Melanoma e1684. Left : Comparaison of the fitted population survival function between groups. Right : Comparaison of the fitted survival function of the susceptible patients between groups.}}
\label{Comparaison}
\end{center}
\end{figure}
\begin{figure}[H] 
\begin{center}
\begin{tabular}{cc}
{
\includegraphics[width=5.35cm,height=5.35cm]{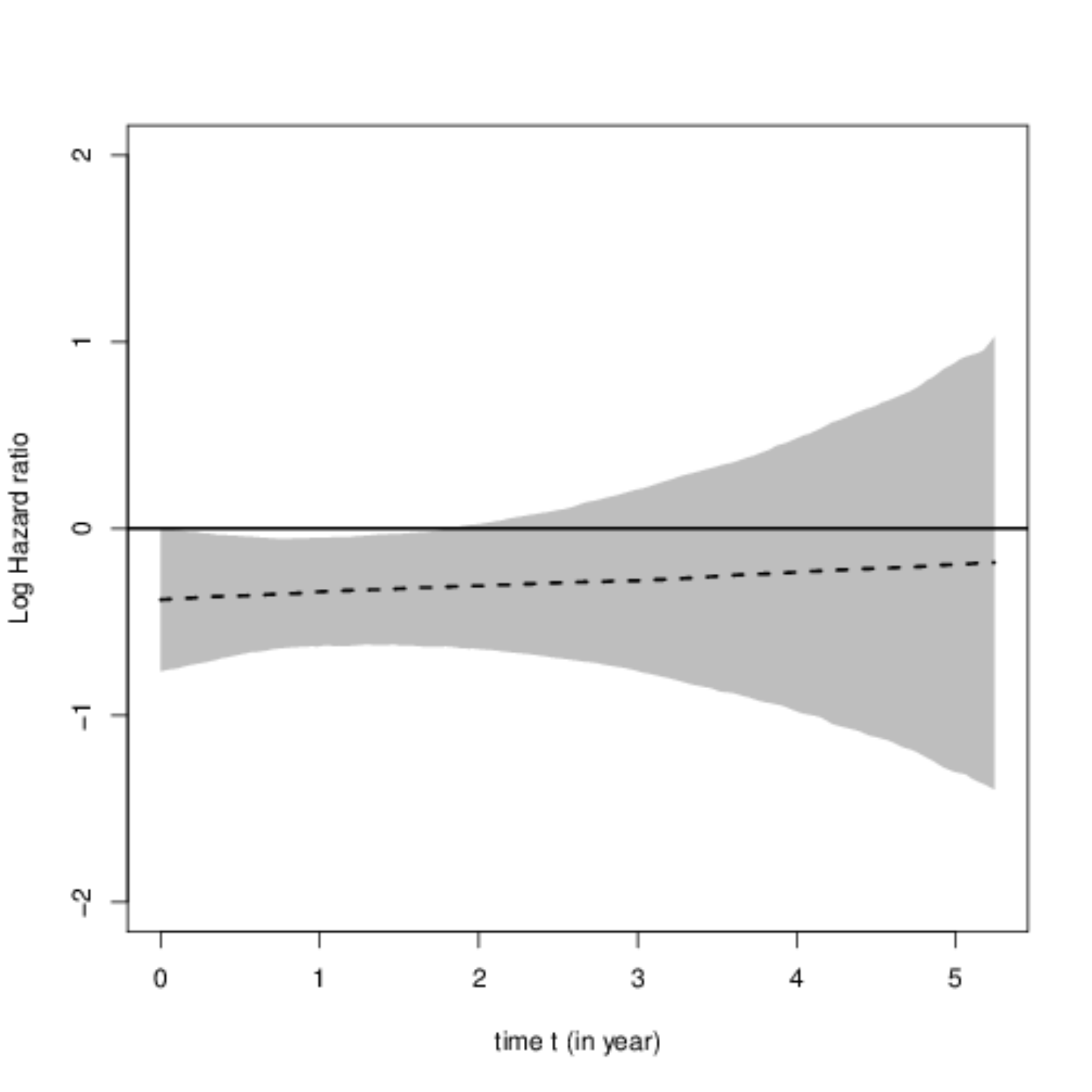} 
}
&
{
\includegraphics[width=5.35cm,height=5.35cm]{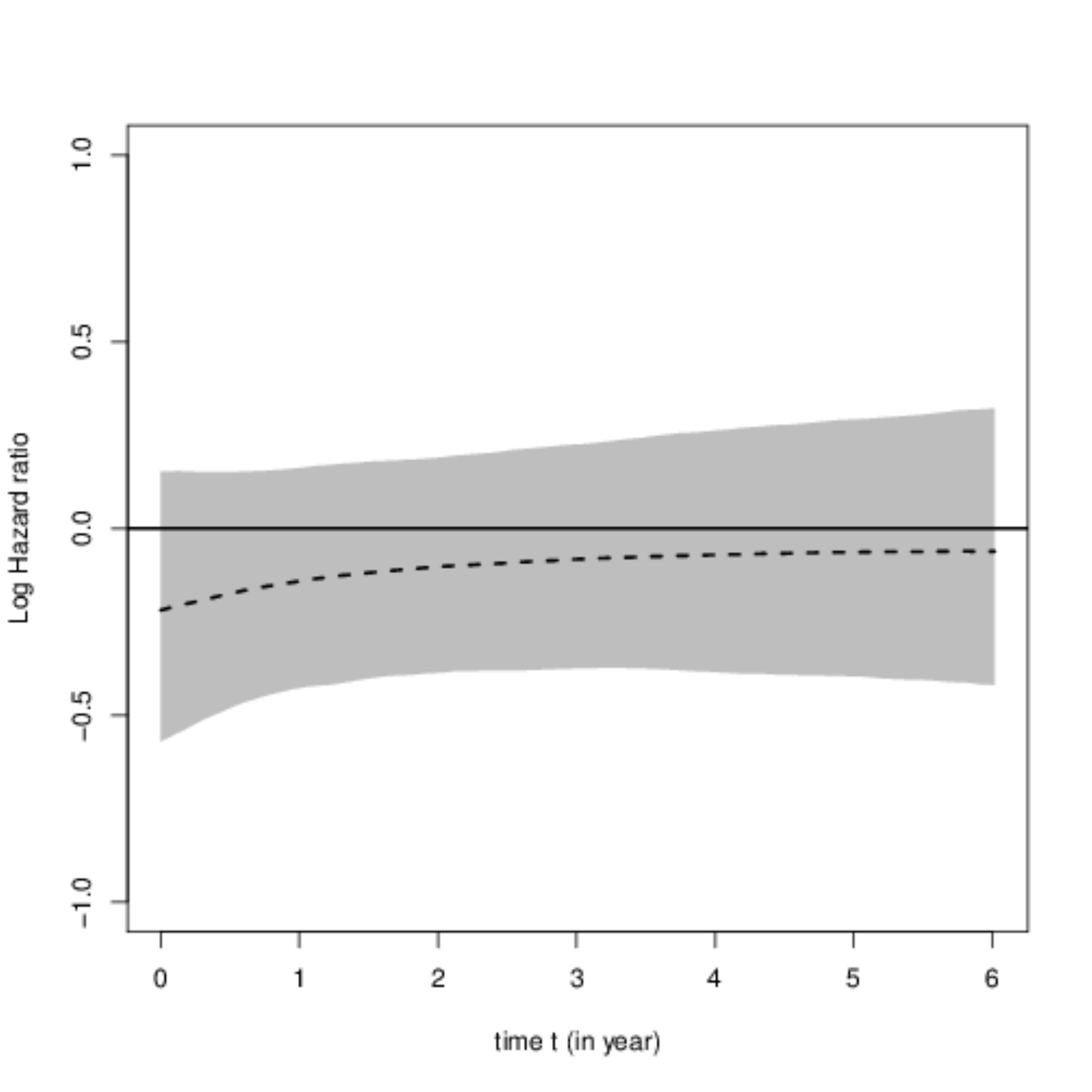} 
}
\end{tabular}
\caption{\footnotesize{Melanoma e1684. Left : Fitted male population log-hazard ratio with $95\%$ pointwize credible region. Right : Fitted log-hazard ratio for the male susceptible patients with $95\%$ pointwize credible region.}}
\label{LOGHR}
\end{center}
\end{figure}

\noindent
We illustrate our methodology on data from a phase III Melanoma e1684 clinical trial. The study was conducted by Eastern Cooperative Oncology Group (ECOG) and consist on a two stages randomized clinical trial : Interferon alpha-2b (IFN) versus Control (Kirkwood et al. (1996)). Their study suggests that Interferon alpha-2b has a significant positive effect on the relapse free survival time. 
Among the $284$ patients present in the study, 144 (51\%) receive the IFN treatment, $171$ $(60\%)$ are male, and we observe a relaspe of cancer for $196$ $(69\%)$ of them. The other subjects were right censored. The age of each subject is known with an average of 47 years and standard deviation of 13 years. The Kaplan Meier estimated curve (Figure \ref{Bremhorst:KMpopulation}) shows a plateau. This suggests that the follow up of the study was sufficiently long. Thus, we let sex, age and the randomised treatment influence simultaneously the probability to be cured and the time necessary for a cell to yield a detectable tumor. \\
The survival data of this clinical trial were already studied in Chen et al (1999) and in Cooner et al (2007).  However, in their analysis, they assumed that the covariates only infuence the probability to be cured. Here, we also enable the covariates to influence the time necessary for a cell to yield a detectable tumor.  \\
The procedure described in Section \ref{PosteriorMCMC}  was use to explore the joint posterior distribution. As for the simulations, $23000$ iterations (including a burnin of $3000$) were generated. To check the convergence of the MCMC algorithm, the traces of all model parameters were examinated critically. The z-scores of the Geweke diagnostics were found to be between $-1.96$ and $1.96$ for all model parameters, suggesting convergence of the MCMC algorithm. \\
Table \ref{Bremhorst:ApplicationCov} presents the MCMC estimates of the posterior median, the $95\%$ HPD interval and the posterior standard deviation of the regression parameters. One can conclude that treatment only has a significant effect on the probability to be cured. In other words, our model suggests that Interferon alpha-2b significantly reduces the number of carcinogenic cells but does not influence the incubation time of one cell. This conclusion is illustrated on Figures \ref{Comparaison} and \ref{LOGHR}. Figure \ref{Comparaison} shows the fitted population survival function (right) and the fitted survival function for the susceptible individuals (left). The only relevant difference is between treatment groups in the population survival function, illustrated for a median value of AGE. Figure \ref{LOGHR}, for males of median age,  shows the estimate of the logarithm of the population hazard ratio $\log(HR_p)$ (left) and of the logarithm of the hazard ratio for the susceptible $\log(HR_u)$ (right) : only $\log(HR_p)$ significantly differs from zero at the beginning of the study. Similar conclusions can be drawn for females.

\begin{table}
\begin{center}
\caption{Melanoma e1684. Estimation of the posterior median, the $95\%$ HPD interval and the posterior standard deviation of each regression parameter of the model.}
\label{Bremhorst:ApplicationCov}
\begin{tabular}{ccccc}
\hline
& Parameters & Estimation & \multicolumn{1}{c}{$HPD_{95\%}$} & $sd_{post}$ \\
\hline
$\multirow{4}{*}{$\theta(\mathbf{x})$}$& Intercept&0.351& [0.088 ; 0.564] & 0.123 \\
& AGE & 0.100 & [-0.046 ; 0.248] & 0.074 \\
  & IFN & -0.321 & [-0.635 ; -0.032] & 0.155 \\
  & SEX & -0.031 & [-0.337  ; 0.276]&0.161 \\
  \hline
   $\multirow{3}{*}{$F(t|\mathbf{z})$ }$ & AGE & -0.136 & [-0.303 ; 0.033] & 0.085 \\
  & IFN & -0.060 & [-0.417 ; 0.341] & 0.190 \\
  & SEX & 0.053 & [-0.333  ; 0.430]&0.195 \\
  \hline
\end{tabular}
\end{center}
\end{table}

\section{Discussion}
\label{s:discuss}
\noindent
A flexible version of the promotion time model when the covariates influence simultaneously the probability of being cured and the time necessary for a cell to yield a detectable tumor was proposed. Although the suggested model does not have a proportional hazard structure at the patient level, our specification provides a smooth estimation of the hazard ratios over time. \\
When the follow up of a study is not sufficiently long, one can use the promotion time model with some restrictions. In this context, it has been proved that of the effects of covariates are identifiable if they are not simultaneously used to model the probability to be cured and the time necessary to detect a tumor growing from a cancerous cell.
The use of a logit link (instead of a $\log$ one) to model the probability to be cured was investigated to try to solve that identifiability problem but it was not successful. \\
Lopes and Bolfarine (2012) propose a parametric promotion time model to deal with hierarchical data. We currently work on the extension of the proposed flexible methodology in this context. An extension to interval censored data will also be considered.
\section*{Acknowledgements}
\noindent
The authors acknowledge financial support from IAP research network P7/06 of the Belgian Government (Belgian Science Policy), and from the contract `Projet d'Actions de Recherche Concert\'ees' (ARC) 11/16-039 of the `Communaut\'e fran\c{c}aise de Belgique', granted by the `Acad\'emie universitaire Louvain'. The authors also thank the editor and the referees for their constructive comments and suggestions for improving this manuscript.\vspace*{-8pt}

\appendix
\section{Proof of lemma 1}
\begin{enumerate}
\item [ ]
\item[A ] Proof of 1). 
\begin{enumerate}
\item [a) ]  Let $(\beta_0, \mathbf{\beta}, \mathbf{\gamma}, S_0)$ and $(\tilde{\beta_0}, \tilde{\mathbf{\beta}}, \tilde{\mathbf{\gamma}}, \tilde{S}_0)$ be two sets of parameters that satisfy (\ref{Bremhorst:modbothcov}) and let $\mathcal{X}$ be the set of all values of vector $\mathbf{x}$. We need to show that if
\begin{eqnarray*}
S_p(t | \mathbf{x}) &=& \exp\left[-\exp(\beta_0 + \mathbf{x}^T \mathbf{\beta}) \left(1 - S_0(t)^{\exp(\mathbf{z}^T \mathbf{\gamma})}\right)\right] \\
&=& \exp\left[-\exp(\tilde{\beta_0} + \mathbf{x}^T \tilde{\mathbf{\beta}}) \left(1 - \tilde{S}_0(t)^{\exp(\mathbf{z}^T \tilde{\mathbf{\gamma}})}\right)\right] \\
&=& \tilde{S}_p(t | \mathbf{x}) \quad \forall \mathbf{x} \in \mathcal{X} \quad ; \quad \forall t \in [0, \infty],
\end{eqnarray*}
then $\beta_0 = \tilde{\beta_0}$ , $\mathbf{\beta} = \tilde{\mathbf{\beta}}$ , $\mathbf{\gamma} = \tilde{\mathbf{\gamma}}$ and $S_0(t) = \tilde{S}_0(t) \quad \forall t \in [0, \infty]$. \\
\item[b) ]Since the exponential function is bijective, we only need to show that if $\forall \boldsymbol{x} \in \mathcal{X} \text{ and } \forall t \in [0, \infty]$, we have 
\begin{eqnarray*}
\exp(\beta_0 + \mathbf{x}^T \mathbf{\beta}) \left(1 - S_0(t)^{\exp(\mathbf{z}^T \mathbf{\gamma})}\right) &=& \exp(\tilde{\beta_0} + \mathbf{x}^T \tilde{\mathbf{\beta}}) \left(1 - \tilde{S}_0(t)^{\exp(\mathbf{z}^T \tilde{\mathbf{\gamma}})}\right) ,
\end{eqnarray*}
then $\beta_0 = \tilde{\beta}_0$ , $\mathbf{\beta} = \tilde{\mathbf{\beta}}$ , $\mathbf{\gamma} = \tilde{\mathbf{\gamma}}$ and $S_0(t) = \tilde{S}_0(t) \quad \forall t \in [0, \infty]$. \\
\item[c) ] Since $F_0(t)$ is a proper cumulative distribution function (see A3), it follows immediately that $F(t|\mathbf{z}) = 1 - S_0(t)^{\exp(\mathbf{z}^T \mathbf{\gamma})}$ is also a proper cumulative distribution function. Let us proof that b) is true for $t = \infty$. Knowing that $F(t|\mathbf{z})$ and $\tilde{F}(t|\mathbf{z})$ are proper cumulative distribution functions, we have to show that if 
\begin{eqnarray*}
\beta_0 + \mathbf{x}^T \mathbf{\beta} &=& \tilde{\beta_0} + \mathbf{x}^T \tilde{\mathbf{\beta}} \quad \forall \mathbf{x} \in \mathcal{X},
\end{eqnarray*}
then $\beta_0 = \tilde{\beta}_0$ and $\mathbf{\beta} = \tilde{\mathbf{\beta}}$. \\
Under A2, this is a straightforward consequence of : 
\begin{eqnarray*}
P_\mathcal{X}(\mathbf{x}^T\mathbf{\beta} = \alpha) = 1 \Rightarrow \alpha = 0 \text{ and } \mathbf{\beta} = \mathbf{0}. 
\end{eqnarray*}
\item[d) ] Since vector $\mathbf{z}$ does not include an intercept (see A1) and under A2, it is well known that the Cox proportional hazard model is identifiable. \\
\item[e) ] Combining the results found in c) and d), we conclude the proof of 1). \\
\end{enumerate}
\item[B) ] Proof of 2). 
\begin{enumerate}
\item[a) ] The follow-up will be said unsufficiently long if the tumor growing from a cancerous cell cannot be detected by the end of the study at time t. It happens if $F_0(t)$ is too small (i.e. close to 0). Then, 
\begin{eqnarray*}
\label{approx}
F(t|\mathbf{z}) &=& 1- S_0(t)^{\exp(\mathbf{z}^T \mathbf{\gamma})} \\
&=& 1 - \left[ \left(1 - F_0(t)\right)^{\exp(\mathbf{z}^T \mathbf{\gamma})} \right] \\
&=& 1 - \left[ 1 - \exp\left(\mathbf{z}^T \mathbf{\gamma}\right)F_0(t) \right ] + o\left(F_0(t)\right)\\
& \approx& 1 - \left[ 1 - \exp\left(\mathbf{z}^T \mathbf{\gamma}\right)F_0(t) \right ] \\
& = &\exp\left(\mathbf{z}^T \mathbf{\gamma}\right)F_0(t).
\end{eqnarray*}
\item[b) ] Thus, using this approximation,  (\ref{Bremhorst:modbothcov}) becomes
\begin{eqnarray*}
S_p(t | \mathbf{x}, \mathbf{z}) & \approx &\exp\left[ -\theta(\mathbf{x}) \exp\left(\mathbf{z}^T \mathbf{\gamma}\right)F_0(t)\right] \\
&=& \exp\left[ -\exp(\beta_0 + \mathbf{x}^T \mathbf{\beta}) \exp\left(\mathbf{z}^T \mathbf{\gamma}\right)F_0(t)\right] \\
&=& \exp\left[ -\exp(\beta_0 + \mathbf{x}^T \mathbf{\beta} + \mathbf{z}^T \mathbf{\gamma})F_0(t)\right].
\end{eqnarray*}
Then, we conclude that if vectors $\mathbf{x}$ and $\mathbf{z}$ share some components, the estimations of the effects of covariates are not identifiable when the follow-up is not sufficiently long. \\
\item[c) ] It remains to prove that the estimations of the effects of covariates are identifiable when the follow up of the study is not sufficienly long and when vectors $\mathbf{x}$ and $\mathbf{z}$ do not share a single component. \\ 
To ensure A3, it is custom to force the \textit{zero tail constraint} : One assumes $\hat{S}_0(t)$ to be 0 beyond the last event time $t_{max}$ (Taylor (1995), Zeng et al. (2006), Ma and Yin (2008)). When the sufficient follow up assumption is not satisfied, this constraint is strong and has some consequences : 
\begin{eqnarray}
\label{Bremhorst::NSLintercept}
\theta(\mathbf{x}) F(t|\mathbf{z})_{|_{t_{max}}} &\approx& \exp(\beta_0 +\mathbf{x}^T \mathbf{\beta}+ \mathbf{z}^T \mathbf{\gamma})  F_0(t)_{|_{t_{max}}} \nonumber \\
&= &\exp(\beta_0 +\mathbf{x}^T \mathbf{\beta} + \mathbf{z}^T \mathbf{\gamma} + \log(1-\epsilon))  \\
&= &\exp(\tilde{\beta}_0 +\mathbf{x}^T \mathbf{\beta} +\mathbf{z}^T \mathbf{\gamma}), \nonumber
\end{eqnarray}
where $\epsilon = F_0(+\infty) - F_0(t_{max})$. \\
In conclusion, the estimations of the covariate effects on the cure probability and on failure time for a cancerous cell are not affected by the zero tail constraint and are thus identifiable if vectors $\mathbf{x}$ and $\mathbf{z}$ do not share some components. 
\end{enumerate}
\end{enumerate}
This concludes the proof of lemma 1.







\end{document}